\documentclass[10pt,twocolumn,aps,pra]{revtex4-2}

\usepackage{xcolor}
\usepackage{microtype}
\usepackage[colorlinks=true,
						linkcolor=blue,
						urlcolor=blue,
						citecolor=blue,
						bookmarks=true,
						pdfborder={0 0 0}]{hyperref}
                        
\usepackage{graphicx}
\usepackage{dcolumn}
\usepackage{bm}
\usepackage{caption}
\usepackage[normalem]{ulem}

\usepackage{siunitx} 
\usepackage{enumerate} 
\usepackage{tikz,pgfplots}
\usepackage{amsmath} 

\usepackage{amssymb}

\usepackage{makecell}

\usepackage{microtype}

\usepackage{comment}
\usepackage{array}
\pgfplotsset{compat=1.18}

\begin{document}

\title{Endogenous Feedback in Coevolutionary Games Reshapes the Stability of Cooperation}
\author{Federico Maria Quetti}

\affiliation{Dipartimento di Matematica, Universit\`a di Pavia, Via Ferrata 5, 27100 Pavia, Italy}

\author{Andrea Civilini}
\affiliation{Sorbonne Université, Paris Brain Institute (ICM), CNRS UMR7225, INRIA Paris, INSERM U1127, Hôpital de la Pitié Salpêtrière, AP-HP, Paris 75013, France}
\affiliation{School of Mathematical Sciences, Queen Mary University of London, London E1 4NS, United Kingdom}
\affiliation{Dipartimento di Fisica ed Astronomia, Universit\`a di Catania and INFN, Catania I-95123, Italy}

\author{Giacomo Frigerio}
\affiliation{Dipartimento di Fisica ed Astronomia, Universit\`a di Catania and INFN, Catania I-95123, Italy}
\affiliation{Dipartimento di Fisica, Universit\`a di Pavia, Via Bassi 6, 27100 Pavia, Italy}

\author{Silvia Figini}
\affiliation{Dipartimento di Scienze Politiche e Sociali, Universit\`a di Pavia, 27100 Pavia, Italy}

\author{Giacomo Livan}
\affiliation{Dipartimento di Fisica, Universit\`a di Pavia, Via Bassi 6, 27100 Pavia, Italy}
\affiliation{Department of Computer Science, University College London, 66-72 Gower Street, WC1E 6EA London, United Kingdom}
\affiliation{INFN, Sezione di Pavia, Via Bassi 6, 27100 Pavia, Italy}

\author{Vito Latora}
\email{v.latora@qmul.ac.uk}
\affiliation{School of Mathematical Sciences, Queen Mary University of London, London E1 4NS, United Kingdom}
\affiliation{Dipartimento di Fisica ed Astronomia, Universit\`a di Catania and INFN, Catania I-95123, Italy}
\affiliation{Complexity Science Hub Vienna, A-1080 Vienna, Austria}

\begin{abstract}


In Evolutionary game theory the payoffs are typically fixed or shaped by external environmental variables. 
Here, we introduce an endogenous-feedback model in which the game played coevolves directly with the population state: the payoff matrix is a time-dependent function of the level of cooperation. This allows strategic incentives to be continuously modified by the collective behavior they generate. 
Even in the simplest case of linear and instantaneous feedback, the model
reveals feedback-induced regimes, termed chimera games, in which stable cooperation arises despite being incompatible with the predictions of standard fixed-game dynamics. 
We further show that delayed feedback can destabilize these equilibria and generate sustained oscillations, while nonlinear feedback reshapes equilibrium structure and introduces path dependence. Our results show how cooperation can be promoted, suppressed, or destabilized by incentives generated endogenously by the very same population's collective behavior. We conclude by outlining how our framework connects to real-world systems shaped by endogenous feedback.

\end{abstract}

\maketitle

\paragraph*{\bf Introduction.}

 Evolutionary game theory (EGT) is the foundational  framework that merges game theory with principles of natural selection. Its rich mathematical structure has been applied extensively to study behavioral dynamics across domains ranging from biology to social and political sciences 
\cite{axelrod1981evolution, maynard1976evolution, taylor1978evolutionary, nowak2006evolutionary}. 
EGT models how behavioral 
traits spread within a population of self-interested agents  \cite{smith1973logic, smith1982evolution, nowak1992evolutionary, weibull_evolutionary_2004}. 
Each agent (player) adopts a strategy and receives a payoff determined by the strategies of the other agents  and  by the specific game setting 
 ~\cite{von2007theory, myerson2013game}.
Strategies then propagate 
through the population according to their fitness, with those yielding higher average payoffs becoming more prevalent.
In biological systems, this mimics the Darwinian principle of survival of the fittest, while in social systems it reflects imitation of the most successful individuals~\cite{livan2019don,de2024imitation}. 
A game is fully specified by its payoff structure, which assigns a payoff to every strategic choice of the players. 
Games of particular relevance, especially for  modeling social environments, are social dilemmas, which capture the tension between individual benefit and collective welfare. 
In such dilemmas, agents choose between two strategies: cooperation, which benefits the group at a personal cost, and defection, which provides a higher payoff when there are cooperators in the population to exploit~\cite{axelrod1988further, doebeli2004evolutionary, nowak2006five, perc2017statistical}. 
Classical examples include  the \textit{Prisoner's Dilemma} and the \textit{Snowdrift Game}  \cite{axelrod1981evolution, maynard1976evolution}.

The game’s payoff structure defines the specific strategic interactions among individuals and, more broadly, the environment in which these interactions occur.
In EGT, the time scales of environmental change and strategic decision-making typically differ: the environment is assumed to be fixed, and only the evolution of strategies within this static setting is considered.
Conversely, real-world systems usually evolve under varying environmental conditions that influence the strategic decisions of interacting individuals~\cite{levin2013social}.
Building on this observation, recent research has begun to extend the evolutionary game framework to incorporate mutable environments, by modeling game settings that change in time with payoffs no longer considered fixed 
\cite{akccay2020deconstructing, chen2025coevolutionary}. 
For example, stochastic games have been used to represent fluctuations in strategy adoption through time-varying payoffs \cite{shapley1953stochastic, wang2014different, hilbe2018evolution, kleshnina2023effect, wang2024evolution}. Yet, these models do not capture the direct effects that population behavior can exert on the environment itself: many real-world systems exhibit \emph{bi-directional feedbacks} in which collective behavior alters the very incentives that subsequently govern strategic competition~\cite{levin2013social,peck1986evolution,bowles2003co,tilman2018revenue,estrela2019environmentally,cortez2020destabilizing}. 
Following this line of reasoning, several co-evolutionary models couple external environmental dynamics to the abundance of strategies in the population \cite{weitz2016oscillating, tilman2020evolutionary, ito2024complete}. 
In such settings, a resource-dependent environment may favor particular strategies, while the resource, and thus the environment itself, is in turn shaped by the strategic composition of the population, creating a feedback loop.
These models, also referred to as eco-evolutionary, are well suited for describing the dynamics of ecological or biological systems driven by reciprocal influences between evolving species and their changing environments, e.g. a rich or depleted resource whose state depends both on its intrinsic regeneration dynamics and on the  cooperative behaviors adopted by agents to preserve it. 

A key finding common to these studies is that feedbacks can qualitatively reshape long-run outcomes, enabling regimes that are absent under fixed incentives, including bistability, sustained oscillations, and stable coexistence~\cite{weitz2016oscillating,tilman2020evolutionary}. In many social and financial settings, however, the relevant “environment” is not a resource stock driven by external renewal or decay, but the incentive structure itself---social norms, expectations, and valuations---which arises endogenously and whose effects manifest exclusively as a consequence of the population's behavior. Positive interactions in online communities, for instance, can make cooperative engagement increasingly rewarding, thereby promoting further cooperation~\cite{borge2013cascading,gonzalez2011dynamics,kang2015rise}. Conversely, in speculative bubbles, collectively sustained expectations of profitability can amplify incentives for opportunistic exit, reinforcing freeriding and abrupt shifts in behavior~\cite{kaizoji2000speculative}. In these cases, strategic interactions do not merely unfold \emph{within} an environment: they continuously \emph{reconstruct} it, implying that the game currently being played may drift or switch endogenously over time. Capturing these phenomena calls for a framework in which the payoff structure is directly coupled to the strategic composition of the population as it dynamically evolves, without any external forcing.

Here, we introduce a general framework for pairwise symmetric co-evolutionary games in 
which the evolution of the game does not rely on the dynamics of external environmental variables, but is endogenously driven by the time-dependent strategy composition of the population. For instance, in the case of social dilemmas, this means that the game played at a given time $t$ depends on the very same level of cooperation in the system. Such a feedback can in general be tuned by choosing a suitable linear or  nonlinear function of the density of the cooperators at a previous time $t - \tau$. The delay parameter $\tau$ allows to control the time it takes for 
the game to feel and react to the incentives due to the level of cooperation in the system. 

Our co-evolutionary model, even in the simplest case of a linear feedback in the density of cooperators, exhibits transitions from one social dilemma class to another. By 
comparing the levels of cooperation observed in our model with those predicted by standard evolutionary game dynamics, i.e. in a static setting, we demonstrate the emergence of game regimes featuring seemingly irrational dynamic equilibria.
We name these regimes \textit{chimera games} as they represent hybrid states in which the emerging game supports a level of cooperation that would not be stable if the same game would have been played in a standard EGT setting. 
Furthermore, delayed responses to feedback and nonlinearity of feedback 
have strong impact on the coexistence of strategies in the population. 
We find that a sufficiently long memory of the system to past levels of cooperation produces cyclic behaviors similar to periodic oscillations found in eco-evolutionary models \cite{tilman2020evolutionary, weitz2016oscillating}, or around chimera games. 
Nonlinear feedback, instead, quantitatively modifies the system's fixed points, reshaping cooperation outcomes and the games at equilibrium.
Overall our results reveal how cooperation can be sustained or suppressed by the very incentives it collectively generates, solely through endogenous feedbacks which continuously reconfigure the payoff landscape itself. 
Moreover, such mechanisms are fundamentally different from the canonical routes to cooperation under fixed payoffs (kin selection, direct/indirect reciprocity, network reciprocity, multilevel selection)~\cite{nowak2006five}.

\bigskip
\paragraph*{\bf RESULTS} 

\paragraph{\bf Evolutionary dynamics of social dilemmas.}
Consider a large population of individuals, which can choose between two pure strategies and interact with a randomly chosen opponent through a pairwise symmetric game. 
Focusing on social dilemmas, the two available strategies are cooperation (C) and  defection (D), and the payoffs are described by the following matrix:
\begin{equation}
    \bordermatrix {
& C & D \cr
C \ \ & R & S \cr
D \ \ & T & P \cr} \ , \label{normal_game}
\end{equation}
where $R$ (reward) denotes the payoff obtained by two cooperating individuals, $T$ (temptation) is the payoff received by a defector against a cooperator, $S$ (sucker’s payoff) is the payoff of a cooperator facing a defector, and $P$ (punishment) is the payoff obtained when both individuals defect.
Without loss of generality, this class of games can be reduced to a two-parameter representation \cite{tanimoto2007relationship, wang2015universal, ito2018scaling}. 
More precisely, the game matrix can be rewritten as:
\begin{equation}
G(D_g,D_r) = 
\begin{pmatrix}
1 & -D_r \\
1+D_g & 0
\end{pmatrix} \ ,
\label{eq:dgdrvalues}
\end{equation}
where $D_g := T-R$ and $D_r := P-S$. These parameters, referred to as \emph{dilemma strength}, define the incentive for an individual to defect in the face of a cooperator or a defector, respectively.
As such, different combinations of $D_g,\ D_r \in [-1,1]$ encompass different classes of games, represented in the space of the two dilemma strengths as a square where each quadrant defines a different class of social dilemma~\cite{pena2016, sadekar2024evolutionary}.
Cooperation equilibria for each of the four social dilemmas are determined by classical results in EGT. In particular, a stable equilibrium $\rho^\star$ of the evolutionary dynamics for a game $G(D_g, D_r)$ coincides with a Nash equilibrium of the game \cite{weibull_evolutionary_2004, cressman_2014_RE}. Fig.~\ref{fig:Classic_game_equilibria} represents the four social dilemma classes and describes their EGT equilibria.

\begin{table}[ht]
\captionsetup{type=figure}
\centering

\begin{minipage}{0.44\linewidth}
\centering
\includegraphics[width=\linewidth]{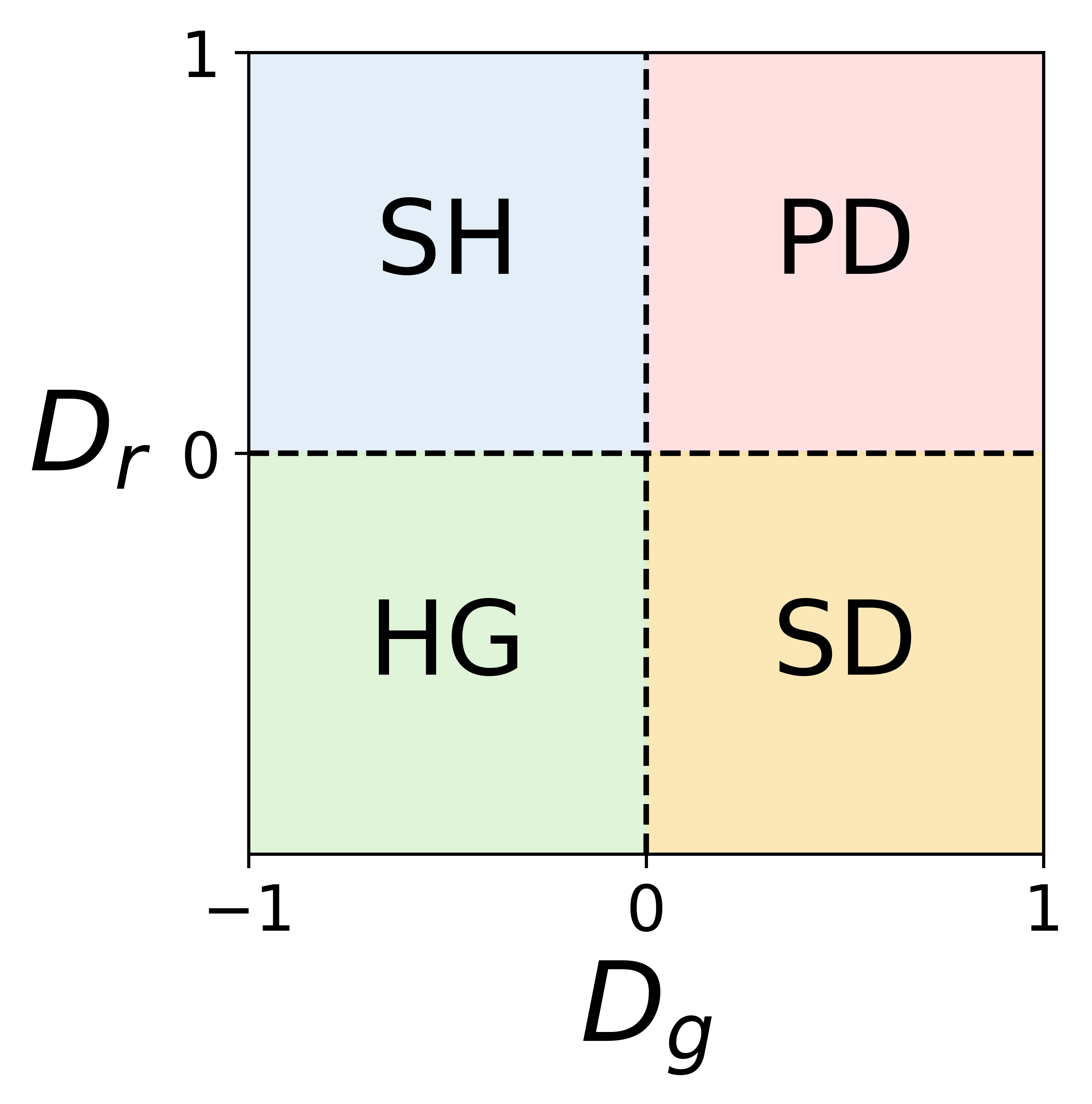}
\end{minipage}
\hfill
\begin{minipage}{0.54\linewidth}
\centering
\scalebox{0.9}{
\renewcommand{\arraystretch}{1.9}
\begin{tabular}{|c|c|}
\hline
\makecell{\textbf{Social} \\ \textbf{Dilemma}} & \textbf{Equilibrium} \\
\hline
\makecell{Prisoner's \\ Dilemma (PD)} & \makecell{Full defection: \\ $\rho^\star = 0$} \\
\hline
\makecell{Snowdrift \\ Game (SD)} & \makecell{Mixed: \\ $\rho^\star \in (0,1)$} \\
\hline
\makecell{Harmony \\ Game (HG)} & \makecell{Full cooperation: \\ $\rho^\star = 1$} \\
\hline
\makecell{Stag \\ Hunt (SH)} & \makecell{Bistability: \\$\rho^\star = 0 \ \mathrm{or} \ 1$} \\
\hline
\end{tabular}
}
\end{minipage}

\caption{Parameter space $(D_g,D_r)$ of the four classes of pairwise social dilemmas, and their EGT equilibria.}
\label{fig:Classic_game_equilibria}

\end{table}

In the literature of EGT, several models add to the evolutionary dynamics of the population a further dynamical variable, inducing changes in the payoff matrix of the game \cite{weitz2016oscillating, tilman2020evolutionary, ito2024complete}. 
The starting point 
of the so-called eco-evolutionary  (ECO-EVO) models is to consider the game $G$ as a convex combination of two different games, $G_1$ and $G_2$, often belonging to two distinct social dilemma classes:
\begin{equation}
\label{eq:endogenus_linear}
 G(n(t)) = (1-n(t))G_1 + n(t)G_2.
\end{equation}
In the equation above, the time dependent weight $n \in [0,1]$ represents the state of the environment, whose evolution is governed by an exogenous dynamical system, mimicking real-world dynamics of depletion/repletion of resources.
A co-evolving model of strategies and environment is then obtained by   
considering a dynamics for $n(t)$ that explicitly accounts for the influence that the population's strategy composition has, in turn, on the resource variable. 
For instance, resource–consumer models by Tilman \emph{et al.}  
\cite{tilman2020evolutionary} 
consider the following dependence of the environmental resource on population’s behavior: cooperative strategies favor regeneration, while defection accelerates depletion, generating a feedback loop between environmental conditions and strategic incentives.
The resulting dynamical systems are bi-dimensional, and their analysis underpins the interplay between the population's strategy composition and the environmental resource, which determines the game, as two coevolving variables.
\begin{figure*}[t]
    \centering
    \includegraphics[width=0.8\textwidth]{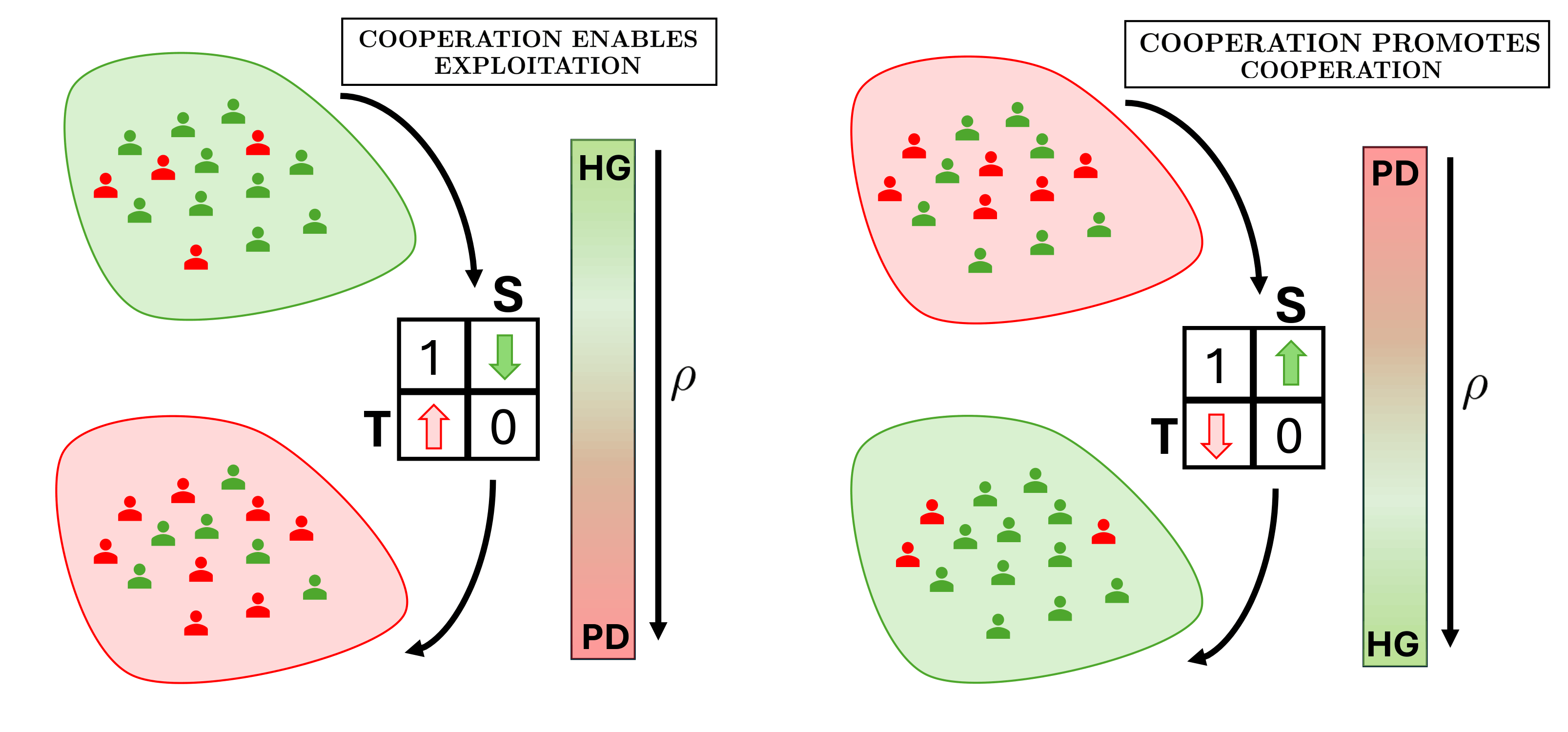}
    \caption{\textbf{Endogenous feedback makes the game depend on the population state.}
    Schematic illustration of the Endogenous-Feedback Model (EFM) proposed in this work. 
    The payoff matrix is not fixed, but changes as a function of the cooperation level $\rho$. This  
    generates a feedback loop between collective behavior and the game being played, so that the social dilemma coevolves with the population composition.
    In the left panel, increasing cooperation makes unilateral defection more rewarding and cooperation more vulnerable to exploitation: the game $G(t)$ is progressively shifted from a Harmony Game toward a Prisoner's Dilemma. 
    In the right panel, increasing cooperation instead promotes mutually beneficial interactions: the game $G(t)$ moves in the opposite direction, from a Prisoner's Dilemma toward a Harmony Game. 
   }
    \label{fig:endogenous_feedback}
\end{figure*}
\\

\paragraph{\bf The Endogenous Feedback Model (EFM).}

Departing from previous models, in which the strategies played by the population and the game are described by two evolving variables, under mutual influence, the co-evolutionary model we propose in this paper does not rely on exogenous dynamics to drive changes in the game. Rather, the game's evolution is assumed to be directly influenced by the level of cooperation $\rho$ in the population at time $t - \tau$, where $\tau$ is a delay parameter.
We thus refer to our model as the Endogenous Feedback Model (EFM) because,
as illustrated in Fig.~\ref{fig:endogenous_feedback}, 
the incentive structure jointly evolves with the strategy distribution of the population only.
The general formulation is specified as follows. The game $G(t)$ played at time $t$ is described by the following payoff matrix:
\begin{equation}
\label{eq:rho_coev}
\begin{split}
 G(t) &\equiv  G(\rho, D^1_g,D^1_r,D^2_g,D^2_r,\tau) \\ &= [1-f\big(\rho(t - \tau)\big)]G_1 +f\big(\rho(t - \tau)\big)G_2 
\end{split}
\end{equation}
where the two games $G_1 = G(D^1_g, D^1_r)$ and $G_2 = G(D^2_g, D^2_r)$ are of the form as in Eq.~(\ref{eq:dgdrvalues}), 
and their convex combination 
is ruled by a function $f:[0,1]\rightarrow[0,1]$ of the level of cooperation in the system, $\rho(t - \tau)$, at the previous time $t-\tau$.
In line with the examples specified in the introduction, decisions on a collective level shape strategic incentives, which in turn shape the population's decision making, in a feedback loop. The delay parameter $\tau$ sets the typical time it takes for the game to respond to the incentives due to the level of cooperation in the system. It plays a similar role to the parameter  controlling the ratio of the time scales of the two processes, namely game evolution and environment dynamics, in standard ECO-EVO models.

By considering large (infinite) unstructured populations, the evolution of the cooperation level $\rho(t)$ 
is ruled by the deterministic Replicator Equation 
\cite{hofbauer2003evolutionary, hofbauer1998evolutionary, taylor1978evolutionary}:
\begin{equation}
    \label{eq:replicator}
\frac{d\rho}{dt} = \rho (1 - \rho) \Delta \pi(\rho) \ ,
\end{equation}
where \( \Delta \pi(\rho) = \pi_C(\rho) - \pi_D(\rho) \) is the expected payoff difference between cooperators and defectors. The payoff difference depends on the strategy frequencies at time $t$ and on the payoffs of the game in Eq.~(\ref{eq:rho_coev}), which depend on the delay $\tau$ (see Supplementary Information for the explicit expression). 
\\
This setting results in different dynamical outcomes from previous coevolutionary models, as the specification of the payoff's evolution is determined only through $\rho$. 
Thus, aside from the different interpretation that our model bears, the resulting one-dimensional dynamics will differ from that of any two-dimensional eco-evolutionary system.
We aim therefore to study the characteristics of dynamical systems arising from our formulation in terms of equilibria and their stability, relating them with the equilibrium game induced through Eq.~\eqref{eq:rho_coev}. 
We compare the equilibrium and stability of cooperation for the game at equilibrium in our EFM formulation with the behavior of a population playing the same game without feedback.
In the main text, we consider two case studies: $G_2$ as a Prisoner's Dilemma and $G_2$ as a Harmony Game, for all possible classes of $G_1$ games, assuming the population starts with initial cooperation level $\rho(t_0) = \rho_0$.  
\\

\begin{figure*}[t]
    \centering\includegraphics[width=0.8\textwidth]{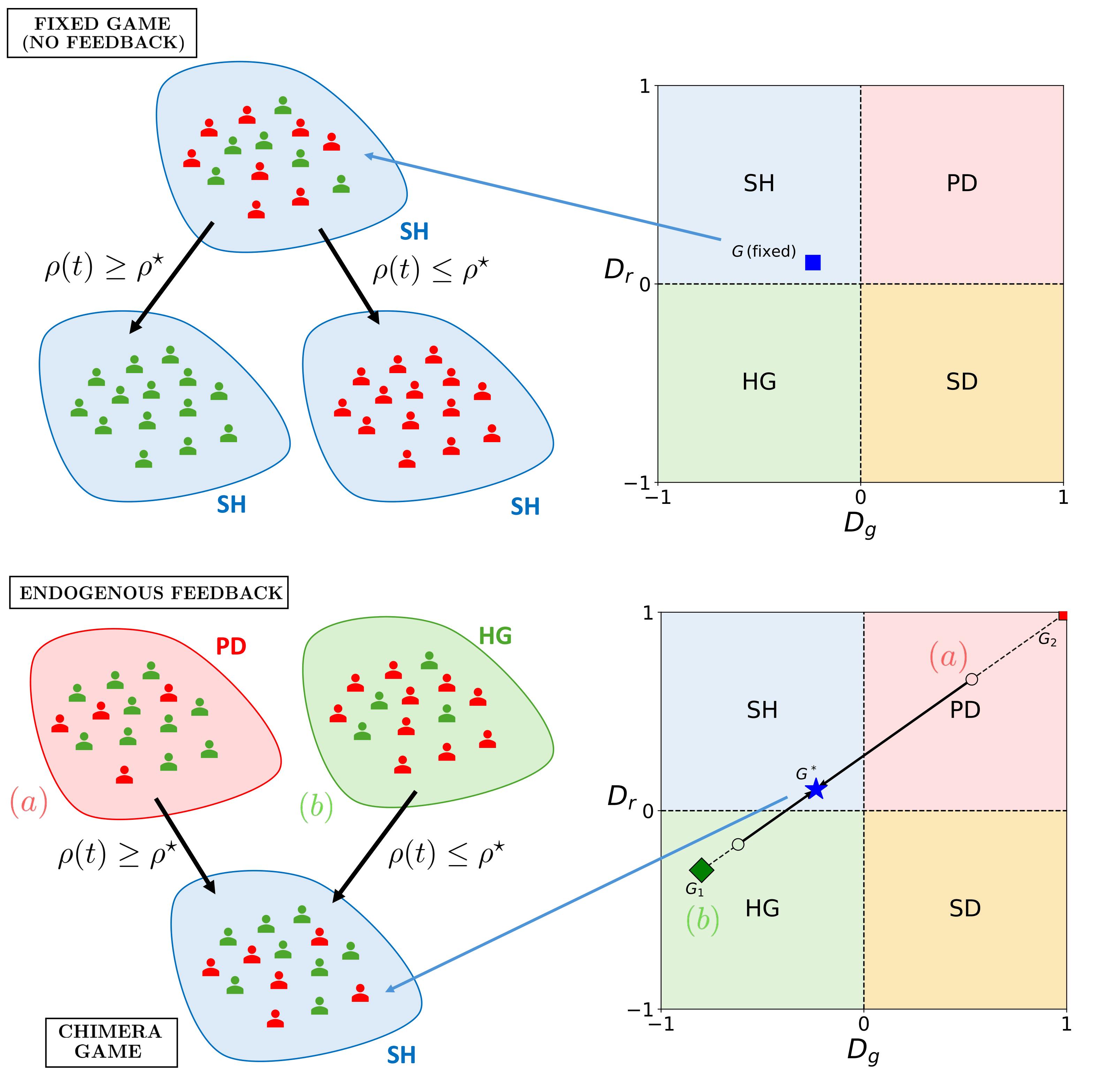}
    \caption{\textbf{Emergence of chimera games in the Endogenous Feedback Model (EFM).}
    Top panels: in the absence of feedback the game played corresponds to a fixed point in the social dilemma plane $(D_g,D_r)$, so the evolutionary outcome is determined by the class of the social dilemma being played.    
    In this specific case, the game being considered is a Stag Hunt, which admits two stable equilibrium states: full cooperation or full defection. 
    Bottom panels: in the EFM the played game $G(t)$ of Eq.\eqref{eq:game} evolves along the segment connecting the games $G_1$ and $G_2$, with the precise position determined by the population cooperation level $\rho(t)$. 
    The insets (a) and (b), connected to the corresponding locations in the Dilemma square, illustrate two representative cases in which distinct underlying dilemmas both drive the dynamics toward the same equilibrium game $G_{eq}$, shown in the Stag Hunt region. 
    We refer to these hybrid regimes as \emph{chimera games}: the population settles at a stable cooperation level satisfying the Nash equilibrium condition, i.e. a Nash equilibrium of the game $G_{eq}$,
    even though such a stable mixed equilibrium would not be supported by the same game under standard dynamics without feedback. In other words, although the equilibrium game lies in the Stag Hunt region, where the static dynamics would allow only full cooperation or full defection (top panel), the endogenous coupling stabilizes an intermediate cooperation level that would otherwise be forbidden.
    Endogenous feedback therefore gives rise to stable strategic outcomes that would be impossible under standard evolutionary dynamics with a fixed payoff matrix.
    }
\label{fig:chimera_games}
\end{figure*}

\paragraph{\textbf{Case study 1: instantaneous feedback ($\tau = 0$).}}
We first consider the simplest case of the EFM, in which the game responds \emph{instantaneously} and \emph{linearly} to the current level of cooperation, i.e. we set 
$\tau = 0$ and $f(\rho)=\rho$ in Eq.~\eqref{eq:rho_coev}. This setting underpins the central mechanism of the model: as $\rho(t)$ evolves, the population does not merely move \emph{within} a fixed game, but continuously reshapes the incentive landscape by shifting the game played at time $t$, namely:
\begin{equation}
\label{eq:game}
G(t)=\bigl(1-\rho(t)\bigr)G_1+\rho(t)G_2,
\end{equation}
which can range along the segment connecting two baseline dilemmas $G_1$ and $G_2$ in the $(D_g,D_r)$ plane (Fig.~\ref{fig:chimera_games}).
Starting from the initial condition $\rho_0$, in the long run the system settles into an equilibrium characterized by a cooperation level $\rho^\star$ and a corresponding \emph{equilibrium game}:
\begin{equation}
G_{\mathrm{eq}}=\bigl(1-\rho^\star\bigr)G_1+\rho^\star G_2.
\end{equation}
As in static social dilemmas, equilibria can be absorbing states, corresponding to consensus of strategies ($\rho^\star=0$ or $\rho^\star=1$), or mixed, corresponding to strategy coexistence ($0<\rho^\star<1$). In light of the dynamic nature of the game in our framework, a key qualitative aspect emerges: whenever stable coexistence is reached, the population typically ends up playing a game $G_{\mathrm{eq}}$ that is \emph{different} from both $G_1$ and $G_2$, and may even fall in a \emph{different} dilemma class than either endpoint.  
The solutions in $(0,1)$ of the dynamical system in Eq.~(\ref{eq:replicator}), determining the stable and unstable levels of cooperation at equilibrium, can be shown (see SI) to correspond to Nash interior equilibria of the game in Eq.~(\ref{eq:game}). 
Importantly, the payoff matrix of this game is itself shaped by the prevailing cooperation level. In this sense, the system evolves over time but eventually settles at cooperation levels that remain consistent with the Nash equilibrium condition. Accordingly, the payoff difference between cooperation and defection vanishes at $\rho^{\star}$.

One of the main findings of our EFM is the following: even a simple linear feedback produces regimes in which \emph{the stability of equilibria} differs from the static interpretation of the equilibrium game.
In classical evolutionary game theory, stability is dictated by the game being played: for instance, a Stag Hunt admits bistability and leads to absorption, whereas a Snowdrift supports stable coexistence (see Fig.~\ref{fig:Classic_game_equilibria}). Under endogenous feedback, however, the system may settle at a stable mixed state even when the corresponding equilibrium game lies in a region whose \emph{static game dynamics} would not allow stable coexistence. We refer to these regimes as \emph{chimera games}: hybrid states in which the population reaches a stable cooperation level that satisfies the Nash equilibrium condition, yet appears inconsistent with the dilemma class of the game being played. In chimera games, equilibria that are unstable under standard evolutionary dynamics without feedback are stabilized by the endogenous coupling, giving rise to qualitatively new dynamical regimes in social dilemmas (Fig.~\ref{fig:chimera_games}).

From this point onward, we focus on two representative choices for the 
game approached as cooperation increases: $G_2=G(D_g^2 = 1, D_r^2= 1)$ (Prisoner’s Dilemma) and $G_2=G(D_g^2 = -1,D_r^2 = -1)$ (Harmony Game). Each setting translates into incentives to different strategies: under increasing cooperation, the former choice dynamically reinforces defective behavior, while the latter rewards further cooperation (see Fig.\ref{fig:endogenous_feedback}).
These two choices generate qualitatively distinct feedback loops and provide the baseline for the delayed and nonlinear cases discussed next. 

\begin{figure*}[ht]
     \begin{minipage}{0.4\linewidth}
    \centering
    \includegraphics[width=\linewidth]{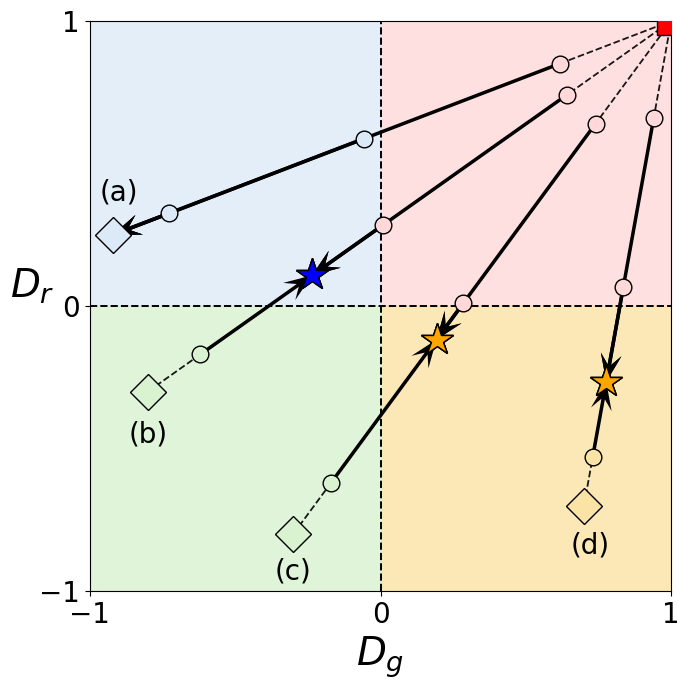}
\end{minipage}
\hfill
\begin{minipage}{0.4\linewidth}
    \centering
    \includegraphics[width=\linewidth]{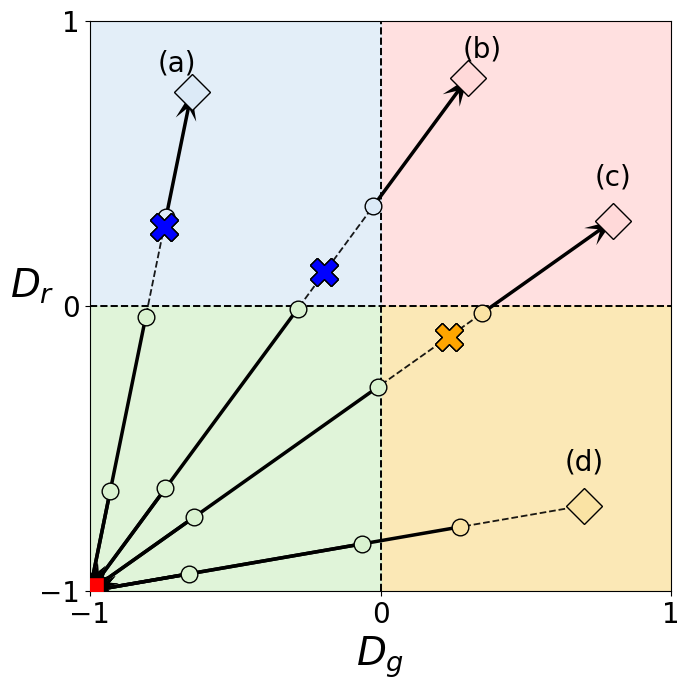}
\end{minipage}

\caption{\textbf{Dynamics of the EFM with linear and instantaneous feedback}.  
Evolution, in the social dilemma square, of various examples of linear and instantaneous EFM games. In each case, the game $G(t)$ played at time $t$ is a time-dependent convex combination of $G_1$ and $G_2$, as in Eq.~\eqref{eq:game}. In the left panel, $G_2$ (red square) is fixed as a PD, and four different $G_1$ games are considered (diamonds).
In the right panel, we adopt the same construction, but fix $G_2$ (red square) as a Harmony Game (HG) instead.
Three initial  cooperation levels $\rho_0 \neq 0$, defining different initial games $G(t_0)$ (circles), are set. 
When the dynamics results in a stable level of cooperation $\rho^{\star}$, the game $G(t)$ converges to non trivial equilibrium games $G_{eq}$, indicated by a star in the social dilemma square.  When the dynamics leads to a non trivial unstable level of cooperation, the corresponding games are represented as a cross.
The equilibrium game $G_{eq}$ in case $(b)$ of the left panel is a  \emph{chimera game}.
}
        \label{fig:game_lambda1}
\end{figure*}

\paragraph{Prisoner's Dilemma: $G_2=G(1,1)$.}
When $G_2$ is a Prisoner’s Dilemma, increasing cooperation shifts the game towards stronger incentives to defect. Despite this 
adverse \lq \lq self-limiting'' feedback, the system can converge to a nontrivial coexistence level $0<\rho^\star<1$ for broad choices of $G_1$ that favor cooperation at low $\rho$ (notably, for $G_1$ in the Harmony or Snowdrift regions). In these cases the equilibrium game $G_\mathrm{eq}$ can lie either in the Snowdrift region, yielding coexistence that matches the static expectation, or in the Stag Hunt region while still sustaining a stable interior equilibrium (see figure \ref{fig:game_lambda1}). 
The latter outcome is precisely a chimera regime: the population converges to a stable mixed equilibrium even though the equilibrium game is a Stag Hunt, which under standard evolutionary dynamics would instead display bistability and absorption.

\paragraph{Harmony Game: $G_2=G(-1,-1)$.}
When $G_2$ is a Harmony Game, increasing cooperation shifts incentives further in favor of cooperation, producing a \lq \lq self-reinforcing'' loop. In this case, the feedback does not support stable interior coexistence: when present, nontrivial fixed points are unstable. The long-run outcome is absorbing, with the final state determined by initial conditions and by whether $G_1$ promotes defection (e.g., PD/SH-like) or cooperation (HG/SD-like). As a result, when $G_2$ lies in the Harmony Game corner, endogenous feedback reshapes basins of attraction but does not generate the stable interior equilibria that, in the Prisoner's Dilemma case, define the chimera regimes.
This linear case study establishes the two key ingredients that will recur in the following sections: (i) equilibrium games can cross dilemma boundaries under endogenous mixing, and (ii) stability can be reshaped by the coupling itself, yielding chimera games even without invoking delay terms or nonlinearities. We fully characterize the behavior under any specification of the pair of games $G_1, G_2$ in the SI. In the next section we show how introducing a finite response time (delay) can destabilize stable coexistence equilibria in the PD-target scenario and induce oscillations.

\bigskip
\paragraph*{\textbf{Case study 2: delayed feedback ($\tau \neq 0$)}} 
In many social and economic systems, incentives respond to collective behavior only after a finite \emph{response time}. Rather than adjusting to the instantaneous population state, institutional and technological mechanisms often rely on observations collected over a finite time window: platform policies are revised only after moderation cycles and monitoring periods, while market valuations may react with delay because of reporting schedules, informational frictions, and bounded attention. In such cases, the relevant ``environment'' for strategic interaction is not an external dynamical variable, but the incentive structure itself, updated on the basis of past rather than instantaneous population statistics.
Within our framework, this finite response time is captured by the delay parameter \(\tau\), so that the  game at time \(t\) depends on the cooperation level observed at the earlier time \(t-\tau\). Delay differential equations have been studied in several contexts within evolutionary dynamics (see e.g.~\cite{alboszta2004stability,mohamadichamgavi2025bifurcation}). More recently, delayed effects have also been considered in models where payoffs are evaluated using past strategy frequencies while the underlying game remains fixed \cite{hu2023evolutionary,hu2026revisiting}. In our EFM, instead, the delayed order parameter determines the \emph{game being played} through endogenous feedback. In the linear case considered below, this corresponds to
\begin{equation}
\label{eq:delay_game}
G(t)=\bigl[1-\rho(t-\tau)\bigr]\,G_1+\rho(t-\tau)\,G_2,
\end{equation}
which is the delayed counterpart of the instantaneous feedback discussed in the previous case study.
Coupling Eq.~\eqref{eq:delay_game} to the replicator dynamics yields a delay differential equation,
\begin{equation}
\label{eq:delay_replicator}
\dot{\rho}(t)=\rho(t)\bigl[1-\rho(t)\bigr]\,
\Delta \pi\!\left(\rho(t),\rho(t-\tau)\right),
\end{equation}
 where $\Delta \pi$ is the payoff difference induced by the game $G(t)$ with delay $\tau$. This term depends on the strategy frequencies at time $t$ as well as at a delayed time $t - \tau$, the latter determining the game matrix (see SI for the explicit expression).
For $\tau=0$ we recover the instantaneous model, which relaxes to fixed points (absorbing or mixed) depending on the games $G_1$, $G_2$. 
For $\tau>0$ the feedback loop can become dynamically self-sustained: beyond a critical delay, when the Hopf condition is satisfied, the fixed point of the instantaneous system loses stability and the dynamics converges to a stable \emph{limit cycle}. 

\begin{figure}[t]
    \centering
    \includegraphics[width=0.99\linewidth]{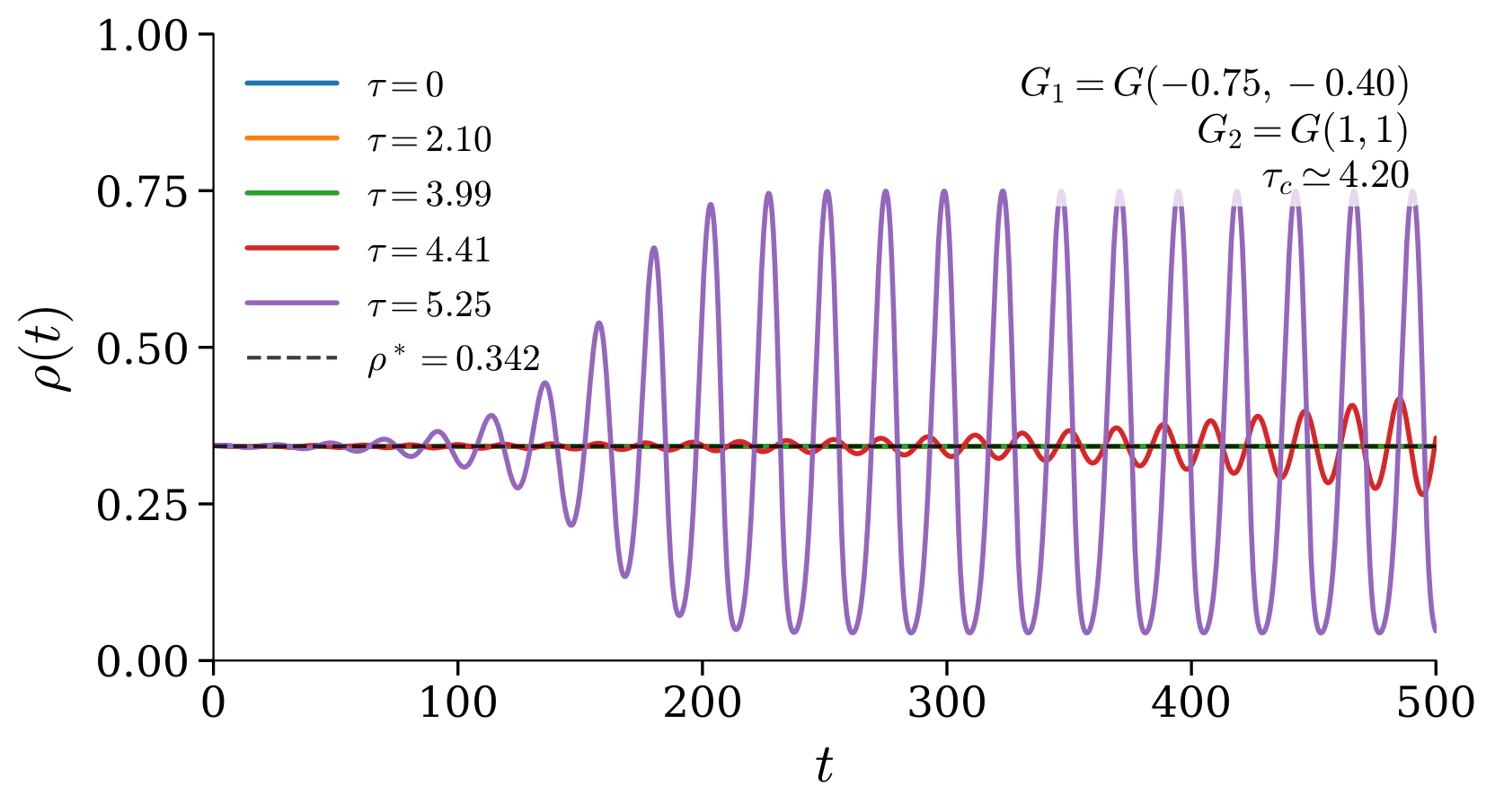}
    \caption{\textbf{Delay-induced oscillations in the EFM with delay.}
    Time series of the cooperation level $\rho(t)$ under delayed incentive feedback for a representative pair of base games
    $G_1: G(D_g,D_r)=G(-0.75,-0.40)$ and $G_2: G(D_g,D_r)= G(1,1)$.
    Curves correspond to different delay values $\tau$ (legend). For small delays the dynamics relaxes to the coexistence equilibrium in the case of instantaneous feedback, whereas for $\tau$ above the analytically predicted threshold $\tau_c\simeq 4.2$ the equilibrium loses stability and the system approaches a stable limit cycle, resulting in sustained oscillations of $\rho(t)$ (see SI for the derivation of $\tau_c$).}
    \label{fig:delay_timeseries_example}
\end{figure}

An illustrative example is shown in Fig.~\ref{fig:delay_timeseries_example}. 
In this oscillatory regime, both the cooperation level $\rho(t)$ and the game $G(t)$ evolve periodically, tracing closed trajectories in the $(D_g,D_r)$ plane. 
Intuitively, as incentives at time $t$ are shaped by earlier behavior, the system persistently \lq \lq overreacts'' to outdated conditions. Given the parameters in the example, cooperation increases under incentives that were generated when cooperation was lower, and decreases under incentives generated when cooperation was higher, producing recurrent boom--bust cycles in behavior and in the incentive landscape.
Since the delay does not affect the location of the fixed points of Eq.~(\ref{eq:replicator}), only altering the behavior of the population in their vicinity,
chimera game scenarios identified in the instantaneous feedback dynamics are preserved. Thus, when chimera games are present, the appearance of delay-induced cycles does not eliminate them, but instead produces sustained oscillations around them. 
Finally, Fig.~\ref{fig:delay_timeseries_example} shows what we demonstrate both analytically and numerically: not only the equilibria but also the stability of the system do not change for insufficiently large delays.

In the following we extend the two configurations already considered in the linear feedback case, namely $G_2=G(1,1)$ (a Prisoner’s Dilemma) and $G_2=G(-1,-1)$ (a Harmony Game). 
The dynamical behavior differs markedly between these two target games. 
A necessary condition for the emergence of oscillatory dynamics is the existence of a stable equilibrium in the instantaneous feedback case (i.e., when $\tau=0$). 
Only in the presence of such a fixed point can the delayed feedback destabilize the equilibrium and give rise to sustained cycles.

\begin{figure*}[t]
\centering
\includegraphics[width=0.99\textwidth]{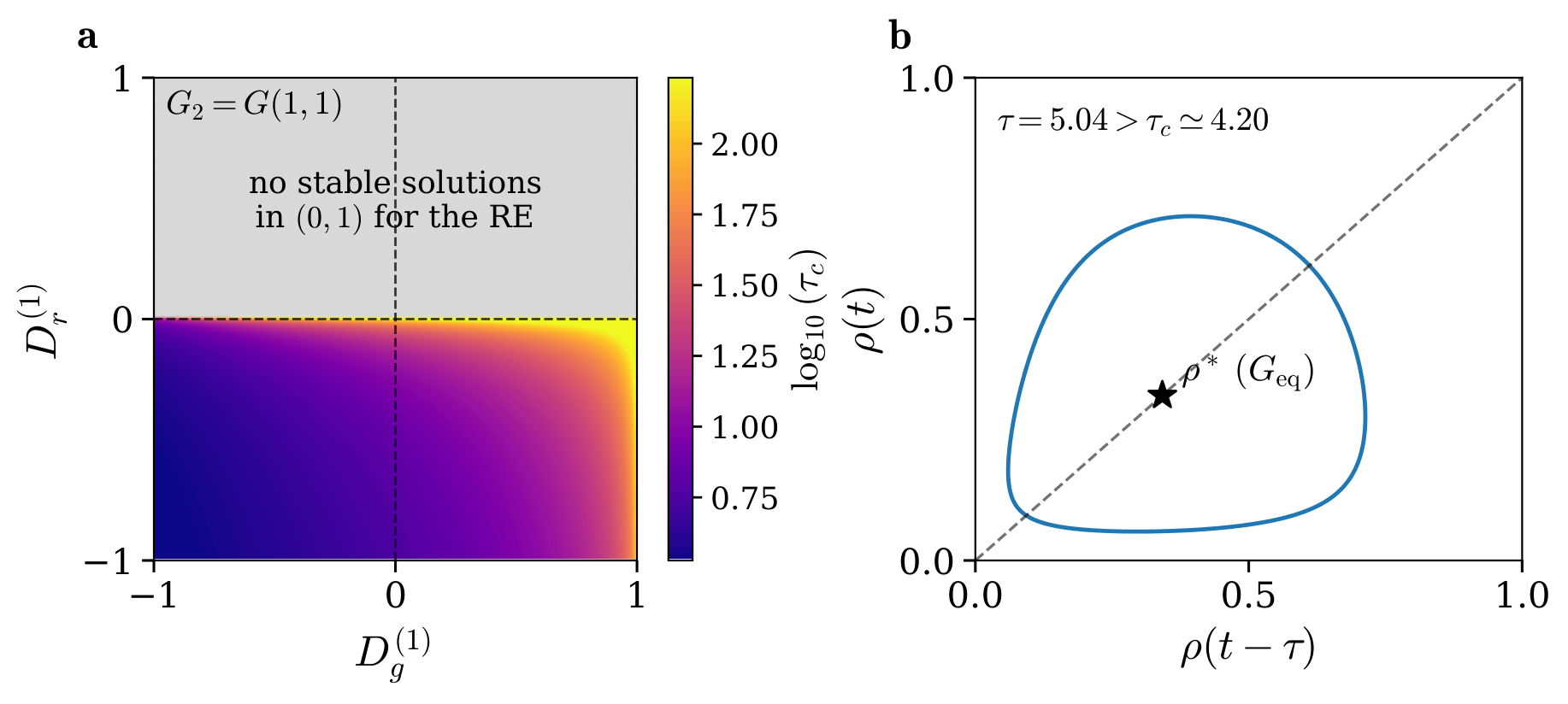}
\caption{\textbf{Delay destabilizes a chimera-game equilibrium and induces sustained oscillations.}
\textbf{a} Critical delay $\tau_c$ for fixed $G_2:G(D_g,D_r)=G(1,1)$, shown as a function of the parameters of $G_1$. 
The threshold value $\tau_c$ is computed from the linear stability analysis of the delayed replicator dynamics around the stable interior equilibrium $\rho^\star$ (see SI), and colours denote $\log_{10}(\tau_c)$. 
The grey region corresponds to parameter values for which the replicator equation admits no stable interior equilibrium in $(0,1)$, and therefore delay cannot induce oscillations through this mechanism. 
\textbf{b} Representation of the cooperation levels at different times, $\rho(t-\tau)$ and $\rho(t)$, for a representative point in the chimera-game regime. 
The blue star identifies the fixed point $(\rho^\star,\rho^\star)$ associated with the equilibrium game $G_{\mathrm{eq}}$. 
This equilibrium belongs to the chimera-game regime, showing that delayed endogenous feedback does not remove the hybrid equilibrium, but can destabilize it and generate oscillatory dynamics around it.}
 \label{fig:tauC_map_G1_G2fixed}
\end{figure*}

As shown in the previous \textit{case study}, when $G_2$ is a Prisoner’s Dilemma the instantaneous dynamics admits stable interior equilibria only for specific choices of $G_1$, namely when $G_1$ lies in the Harmony Game or Snowdrift regions of the dilemma space. 
In these parameter regimes, introducing a delay in the feedback can destabilize the fixed point. 
Indeed, beyond a critical delay $\tau_c$, the equilibrium loses stability through a Hopf bifurcation and the system transitions to a regime of self-sustained oscillations (see SI for the analytical derivation). 
Fig.~\ref{fig:tauC_map_G1_G2fixed} reports the values of the critical delay $\tau_c$ across the regions of the game space where such oscillatory dynamics can occur.
We next consider the alternative configuration in which $G_2=(-1,-1)$, corresponding to a Harmony Game. 
In this case, the instantaneous dynamics admits only absorbing equilibria at $\rho=0$ and $\rho=1$, with no stable interior fixed points. 
As a result, introducing delayed feedback does not generate additional dynamical regimes: the system still converges to the same absorbing states as in the instantaneous model, thus the delayed formulation exhibits the same behavior as the linear feedback case.

\bigskip 
\paragraph*{\textbf{Case study 3: nonlinear feedback.}}

The linear mapping $f(\rho)=\rho$ represents the simplest possible type of endogenous coupling between the population state and the incentive structure. 
In many real-world systems, however, the impact of collective behavior on strategy adoption is unlikely to be strictly linear. 
Institutional responses, social norms, and market incentives often display amplification or saturation effects: the impact of additional cooperative behavior may increase once cooperation becomes widespread, or conversely diminish when the system approaches saturation, depending on the specific scenario.

To capture these effects, we have considered in our EFM the nonlinear feedback of the form
\begin{equation*}
f(\rho)=\rho^{\lambda}, \qquad \lambda>0,
\end{equation*} 
which preserves the proportionality between cooperation levels and games being played: larger values of $\rho$ still correspond to games closer to $G_2$ while $\lambda$ controls how rapidly this transition takes place. 
Values $\lambda>1$ correspond to a slower response of the incentive structure to changes in cooperation, reflecting saturation effects, whereas $\lambda<1$ produces a faster response that amplifies the influence of cooperative behavior.
Finally, for $\lambda=1$ the model reduces to the linear case discussed above.
For nonlinear instantaneous feedback, the game played at time $t$ becomes
\begin{equation}  
G(t) = [1-\rho^{\lambda}(t)]G_1 + \rho^{\lambda}(t) G_2.
\end{equation}
Since the condition for fixed points is determined by the payoff difference $\Delta\pi(\rho)$, introducing the nonlinear response $f(\rho)=\rho^{\lambda}$ modifies how $\Delta\pi(\rho)$ depends on the population state and consequently the conditions for cooperation states and corresponding equilibrium games.  
Increasing $\lambda$ changes the mapping between cooperation and the game being played by making the feedback less sensitive at intermediate values of $\rho$. Indeed, as $0<\rho<1$, one has $\rho^\lambda<\rho$ when $\lambda>1$ (superlinear scenario), so the game played at time $t$ remains closer to $G_1$ unless cooperation is already high. 
This parameter choice leads to a shift of the stable fixed point toward larger cooperation levels. By contrast, when $0<\lambda<1$ (sublinear scenario), one has $\rho^\lambda>\rho$, so the feedback reacts more strongly already at intermediate cooperation levels. Thus, the game shifts more rapidly toward $G_2$ for higher values of $\rho$, generally resulting in a lower equilibrium cooperation level (see SI).
Equivalently, feedback with parameter $\lambda>1$ can be interpreted as a more gradual, attenuated response of incentives to cooperation, whereas $\lambda<1$ as an amplified response. This nonlinear sensitivity is also reflected in the set of equilibrium games reachable by the dynamics: for $\lambda>1$, the slower drift from $G_1$ to $G_2$ enlarges the region of the dilemma plane that can be attained at equilibrium, while for $\lambda<1$ the stronger response compresses that region by driving the system more quickly toward games closer to $G_2$ (see SI).

\begin{figure*}[ht]
     \begin{minipage}{0.24\linewidth}
    \centering
    (A) $G_2: G(1,1); \quad \lambda=0.5$
    \includegraphics[width=\linewidth]{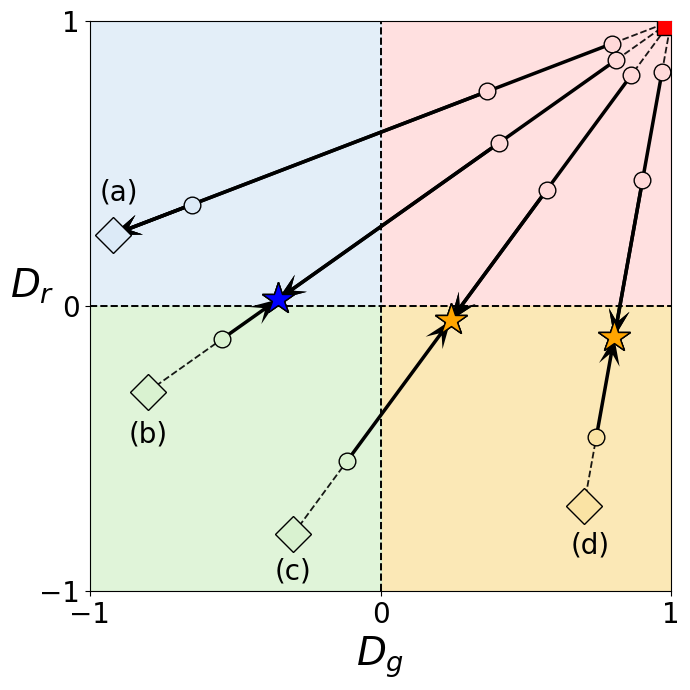}
\end{minipage}
\hfill
\begin{minipage}{0.24\linewidth}
    \centering
    (B) $G_2: G(-1,-1); \ \ \lambda=0.5$
    \includegraphics[width=\linewidth]{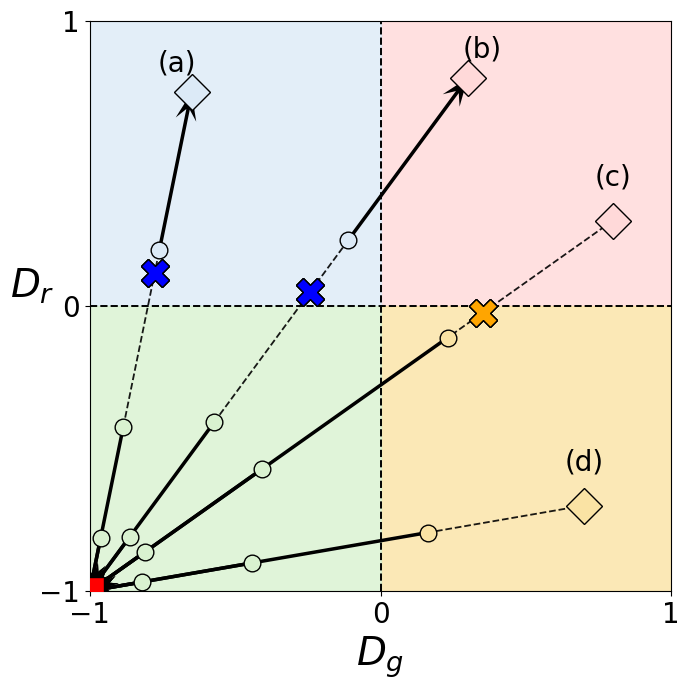}
\end{minipage}
\hfill
\begin{minipage}{0.24\linewidth}
    \centering
    (C) $G_2: G(1,1); \quad \lambda=2$
    \includegraphics[width=\linewidth]{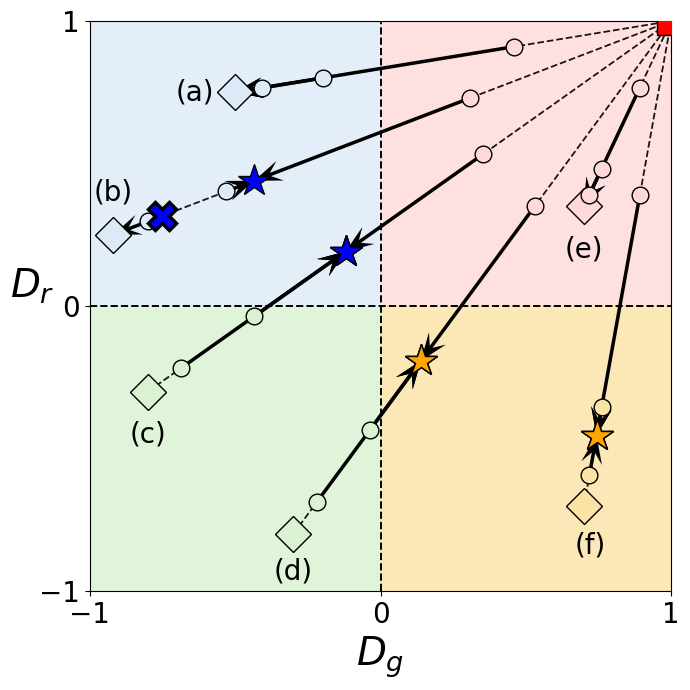}
\end{minipage}
\hfill
\begin{minipage}{0.24\linewidth}
    \centering
    (D) $G_2: G(-1,-1); \ \ \lambda=2$
    \includegraphics[width=\linewidth]{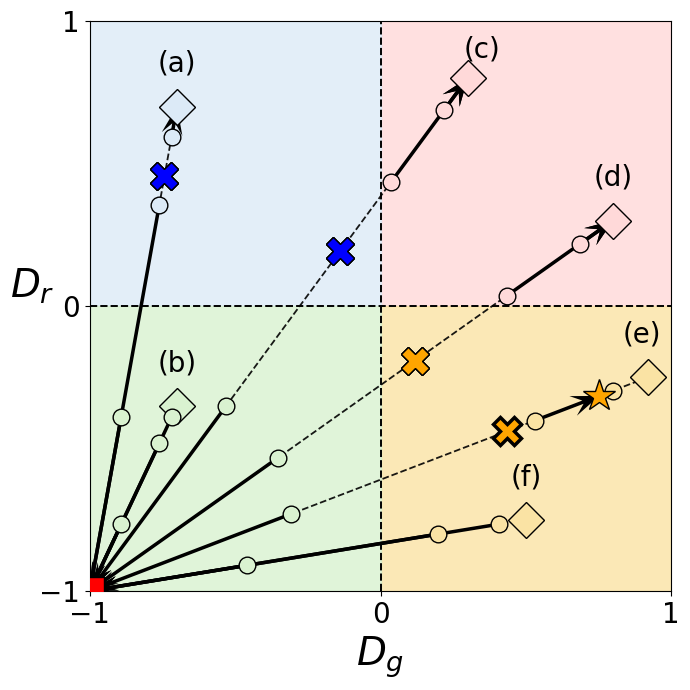}
\end{minipage}

\caption{\textbf{Dynamics of the EFM with nonlinear and instantaneous feedback.}  
The evolution of the game $G(t)$ represented in the social dilemma square. The game $G_2$ is represented in each case as a red square. Four different $G_1$ games are considered (diamonds), and three initial cooperation levels $\rho_0$, defining different initial games (circles), are shown. The game $G(t)$, which is specified as a convex combination of $G_1$ and $G_2$, shifts between them. When the game $G(t)$ converges to a non trivial equilibrium game $G_{eq}$, it is represented by a star, while nontrivial unstable equilibria repel the configuration and are represented as a cross.
The rate of feedback $\lambda$ affects the position of allowed games at equilibrium in the parameter space.
Moreover, while plots (A) and (B) for sublinear feedback resemble the linear results in Fig.\ref{fig:game_lambda1} in terms of equilibria, multiple equilibria emerge for certain $G_1$ choices, as seen in the superlinear scenarios of plots (C), (D).} 
\label{fig:lambda_nonlinear}
\end{figure*}

The effect of a nonlinear mapping is not limited to a shift in the location of dynamical equilibria.
Because $\Delta\pi(\rho)$ now depends nonlinearly on the population state, the replicator dynamics can in principle admit multiple interior fixed points in the interval $\rho\in(0,1)$, reflecting the richer structure induced by nonlinear feedback. 
To illustrate these effects, we consider two representative values of the nonlinear exponent: $\lambda=0.5$, corresponding to sublinear feedback, and $\lambda=2$, corresponding to superlinear feedback. We find that sublinear feedback does not qualitatively alter the outcomes of the two case studies considered here, namely the cases in which $G_2$ is a \textit{PD} or a \textit{HG}, relative to the linear case (see Fig.~\ref{fig:game_lambda1} and Fig.~\ref{fig:lambda_nonlinear}). Depending on the choice of the game $G_1$, the system either admits no coexistence equilibrium or exhibits a single interior equilibrium, which is stable when $G_2$ is a \textit{PD} and unstable when $G_2$ is a \textit{HG}.
By contrast, superlinear feedback can change the set and stability of equilibria in both scenarios. In addition to the cases described above, two interior fixed points with $\rho\in(0,1)$ may emerge, necessarily with opposite stability. As a result, for the same game parameters the system can display bistability between consensus and coexistence, so that the long-run outcome depends on the initial cooperation level $\rho_0$ (see Fig.~\ref{fig:lambda_nonlinear}). This is in contrast with the linear-feedback case, where each choice of game parameters selects a single equilibrium class—either consensus or coexistence—independently of the initial state.

Overall, nonlinear feedbacks in the EFM enrich the outcomes of the linear case: sublinear responses mainly shift equilibria, whereas superlinear responses can generate multiple equilibria and initial-condition dependence. The exponent $\lambda$ thus acts as a key control parameter of the long-run dynamics of cooperation and game played. \\

\paragraph*{\bf Discussion.}
Understanding the emergence of cooperation is a fundamental problem in evolutionary game theory~\cite{axelrod1981evolution, nowak2006five, hofbauer1998evolutionary, nowak2006evolutionary}. 
While standard models of populations whose individuals interact through a social dilemma capture the basic strategic conditions for cooperation, exemplified by Nowak’s five rules for the evolution of cooperation \cite{nowak2006five}, they typically assume fixed environmental conditions.
In many real-world systems, however, the state of the environment is shaped dynamically by the collective behavior of the population, and a number of modeling approaches have begun to incorporate such coupled dynamics~\cite{weitz2016oscillating, tilman2020evolutionary, betz2024evolutionary, perc2010coevolutionary, gross2008adaptive,delton2011evolution, hilbe2018evolution, sadekar2024evolutionary, civilini2024explosive}.

In analogy with this setting, Eco-evolutionary models feature a payoff matrix that is modified through an additional dynamical variable describing the state of the environment~\cite{weitz2016oscillating, tilman2020evolutionary, ito2024complete}. In such frameworks, the game is typically represented as a convex combination of two games, with a weight determined by an exogenous environmental variable whose dynamics account for depletion, regeneration, or other forms of environmental change. The resulting systems describe the coupled evolution of strategy frequencies in the population and the state of the environment as two distinct coevolving variables. In our model, instead, no separate environmental dynamics is invoked: the payoff structure is directly shaped by the fraction of cooperators, so that the feedback is entirely endogenous to the population state and allows the game to range between dilemmas that entail different strategic tensions and cooperation outcomes.
In our analysis, three variants are considered, each describing a different kind of response from the system to changes in its composition: instantaneous linear feedback, delayed  feedback, nonlinear feedback.

Firstly, the case of linear and instantaneous feedback captures situations in which the game responds immediately to the current state of the population. In this setting, an equilibrium level of cooperation is reached together with a corresponding equilibrium game. However, the nature of the equilibrium, either coexistence of strategies or consensus towards one strategy, is determined by the feedback between population behavior and incentives, rather than by a fixed game alone. 
When the feedback is such that increasing cooperation reinforces defection, a coexistence equilibrium can appear. Instead, when cooperation  reinforces itself no internal equilibrium can arise, and the population collapses to an absorbing state, which, depending on the initial conditions, can either be full cooperation or full defection.
As a result, instantaneous endogenous feedback can move the system across social-dilemma classes and generate stability-switching phenomena, including chimera games: feedback-induced regimes in which stable cooperative states arise despite being incompatible with the predictions of standard evolutionary dynamics for the corresponding fixed game. The reverse may also occur, with otherwise stable equilibrium outcomes no longer reached because feedback draws the system toward consensus. \\ 
Our results can have practical applications as many situations can be modeled as systems with endogenous feedback operating on near-instantaneous timescales. An example is provided by highly liquid electronic financial markets, where order flow, liquidity provision, and price changes are incorporated into trading decisions on extremely short timescales \cite{aquilina2022quantifying, brogaard2014high}. In such markets, cooperation-like behavior among traders, such as supplying liquidity or contributing to price discovery, continuously reshapes the incentives faced by other participants. This can generate chimera-like outcomes: liquidity provision may persist even though the created liquidity also strengthens incentives for latency arbitrage, adverse selection, or withdrawal by faster or better-positioned traders. Thus, the observed coexistence of liquidity provision and opportunistic exploitation may appear at odds with a fixed-game interpretation, while becoming more natural once incentives are treated as endogenously shaped by collective market behavior.
Another example that can be naturally modeled within this framework is provided by real-time reputation and feedback systems in online platforms. These systems are usually studied using standard game-theoretic tools, including repeated games, signaling models, and mechanism design \cite{tadelis2016reputation}. Nevertheless, they display a tension that is evocative of chimera games. In anonymous online exchange, such as online Q\&A and sharing economy platforms, the apparent one-shot interaction may seem to favor opportunistic behavior. Yet, cooperative actions such as reliable delivery, honest exchange, or helpful participation are converted into reputational signals that rapidly affect trust, visibility, matching probabilities, and future gains from interaction \cite{bolton2004effective}. Reputation systems therefore exemplify how cooperation levels that might appear puzzling under a fixed-game interpretation can be understood in light of a payoff structure that is allowed to depend on population-level feedback.

The second case we considered, the delayed-feedback case, models scenarios in which the game is affected by past states of the system, rather than its most recent configuration.
In this setting, the mathematical structure of the game is defined as in the instantaneous case, with the only difference that
the feedback is given by a previous level of cooperation in the population, thus describing a delayed response on the game played at time $t$.
As a result, equilibria that would otherwise be stable in the case of instantaneous feedback can become unstable for sufficiently large delays. The system then no longer converges to a fixed cooperation level and a fixed equilibrium game, but instead displays self-sustained oscillations in both the level of cooperation and the game being played. In this regime, feedback can repeatedly move the system across social-dilemma classes, producing cycles between strategic environments that alternately promote and undermine cooperation.

This delayed-feedback mechanism is suggestive of real-world settings in which incentives are shaped by outdated information about collective behavior. One example is provided by relatively opaque or illiquid financial markets, such as corporate or municipal bond markets, where information diffusion is slower than in highly liquid electronic markets and prices may adjust only gradually once relevant information becomes observable \cite{cao2023implied}. In such settings, agents may react to past signals of liquidity, risk, or mispricing, generating delayed adjustments in liquidity provision, trading activity, or market participation. A second example is vaccination behavior. In standard vaccination games, vaccine uptake affects disease prevalence, which in turn modifies the perceived payoff of vaccination, thereby generating oscillations in both uptake and disease prevalence \cite{bauch2005imitation}. Our endogenous delayed-feedback framework can be seen as a stylized abstraction of this loop, omitting the intermediate epidemiological variable and letting incentives respond directly to current or past behavior: it isolates the dynamical consequences of endogenous feedback, including oscillatory cooperation-like dynamics. Similar delayed responses may also arise in platform management \cite{madio2025content}, where moderation policies are revised after observing user behavior and advertiser risk over finite time windows, and in adaptive resource management, where policies are updated using lagged measurements of resource use, depletion, or recovery \cite{walters1986adaptive}.

Finally we have considered the case of instantaneous nonlinear feedback.
Here, as feedback is not linearly proportional, the incentive structure responds more/less strongly to a given level of cooperation, depending on the type of nonlinear mapping between population state and game. Nonlinear feedback can preserve the chimera-game mechanism, while shifting the regions in which chimera equilibria occur. Moreover, superlinear responses can allow coexistence and consensus to arise for the same underlying pair of games. The particular type of outcome obtained depends on the initial level of cooperation, adding path dependence to the stability-switching effects already observed in the linear case. 
By contrast, a sublinear feedback does not modify qualitatively the linear feedback outcomes, as the dynamics converges to a single class of equilibrium regardless of the initial condition. Therefore, different nonlinear response profiles correspond to different sets of admissible cooperative states, reshaping the space of equilibrium games.

The nontrivial coevolution between population strategies and rewards which the nonlinear case describes is naturally connected to social norms, critical mass, and pro-environmental behavior. Cooperative actions such as low-carbon consumption, energy saving, or sustainable mobility may have limited impact when rare, but become increasingly attractive once sufficiently widespread, as social approval, imitation, and perceived efficacy begin to reinforce them \cite{granovetter1978threshold, centola2018experimental, otto2020social, winkelmann2022social}. 
Similar effects arise in technology adoption, where network externalities can generate lock-in or tipping \cite{katz1985network}, and in public-health compliance, where behaviors such as mask wearing, testing, or isolation may become self-reinforcing once socially expected and saturate once widely adopted.
Such threshold effects are evocative of chimera games: cooperation may appear fragile or inconsistent with the apparent static incentives of the system, yet become stable once collective behavior has nonlinearly reshaped the payoff landscape.
\\

Overall, the framework we have introduced and the results obtained demonstrate that the introduction of an endogenous feedback can substantially reshape the behavior of the system, producing cooperation levels that arise from feedback-driven changes in the game being played. As a consequence, the transition away from the initial game can shift the system into a different social dilemma class. Remarkably, we find feedback-induced equilibrium regimes, such as the chimera games, where stable cooperative states evade the predictions of standard evolutionary dynamics. 
These phenomena, which do not arise in game-theoretic frameworks with fixed payoff structures, also have implications for real-world systems.
Policymakers must recognize that dynamic responses to incentives may not merely shift individual behavior but can also disrupt the stability of cooperation by altering the structure of strategic interactions. Such feedback-driven changes can lead to outcomes that static (i.e. no-feedback) payoff-based models are unable to anticipate. Consequently, policy and mechanism designs that operate through dynamic feedback must take into account that, in social, ecological, and economic settings, the environment is not a fixed backdrop but a malleable and coevolving component of the system~\cite{weitz2016oscillating, tilman2020evolutionary, ito2024complete, sadekar2024evolutionary, gross2008adaptive, civilini2024explosive, civilini2021evolutionary, bowles2003co, sadekar2025drivers}. \\

\paragraph*{\bf Methods} 
\paragraph{Analysis.}
We consider a coevolutionary social dilemma in which the game at time $t$ is a state-dependent convex combination of two baseline games, so that incentives feed back on the current or past level of cooperation. The dynamics in well-mixed populations is described by replicator equations for both instantaneous and delayed endogenous feedback. The analytical treatment of equilibria, local stability, chimera-game regimes, and delay-induced oscillatory instabilities is presented in the Supplementary Information, which also contains the explicit parametrization of the game and all derivations.

\paragraph{Simulations.}
We complemented the deterministic analysis with stochastic agent-based simulations in finite well-mixed populations using pairwise comparison dynamics under the same endogenous-feedback rule. In the case of delayed feedback, the game was updated using the cooperation level measured at a previous time. Details of the simulation algorithm, parameter values, numerical implementation, and comparison with the deterministic limit are given in the Supplementary Information. The simulations shown in the main text can be reproduced using the code referenced in the Code availability section.

\paragraph{Code availability} 
The code used for the figures, the MC simulations and the ODE/DDE resolvers is available at \textit{https://github.com/GiacomoFrigerio/CoevChimeraGames}.

\paragraph{Data availability }
All data shown in the figures can be generated using the simulation code provided in the Code Availability section.

\paragraph*{Acknowledgments.}
F.M.Q. acknowledges RES for the award of a Ph.D. scholarship.
A.C. and V.L. acknowledge support from the European Union -  NextGenerationEU, GRINS project (grant E63-C22-0021-20006).

\paragraph{Author contributions}
F.M.Q., A.C. and G.F. contributed equally to this work. \\
F.M.Q.: conceptualization, methodology, simulations, formal analysis, writing (original draft, review, editing). A.C.: conceptualization, methodology, formal analysis, writing (original draft, review, editing), supervision.
G.F.: conceptualization, methodology, simulations, formal analysis, writing (review, editing).
S.F.: supervision.
G.L.: conceptualization, methodology, formal analysis, writing (original draft, review, editing), supervision.
V.L: conceptualization, methodology, formal analysis, writing (original draft, review, editing), supervision.

\paragraph{Competing interests}
The authors declare no competing interests.

\newpage



\clearpage
\onecolumngrid

\renewcommand{\theequation}{S\arabic{equation}}
\renewcommand{\thefigure}{S\arabic{figure}}
\renewcommand{\thetable}{S\arabic{table}}

\setcounter{equation}{0}
\setcounter{figure}{0}
\setcounter{table}{0}

\section*{Supplementary Information}

\section{Simulations}

Here we detail the Monte Carlo simulation protocol performed to validate the analytical results.
The standard evolutionary framework adopted for simulations relies on the well mixed assumption and draws from the replicator equation \cite{szabo1998evolutionary, traulsen2006stochastic
}. 
Consider a finite but sufficiently large population of agents (chosen in our analyses as $N=2500$), sharing a game at full defection determined by the payoff matrix $G_1$. The population is initialized to a cooperation level $\rho_0$, and each agent equipped with an initial strategy at random. This leads in our model to an initial feedback-driven game $G (t = 0) = (1 - \rho_0) G_1 + \rho_0 G_2$. 
Upon specifying the initial game dilemma strengths and the $G_2$ game dilemma strengths, the payoff matrix is completely defined. Note that populations endowed with different $G_1, \rho_0$ but same $(1 - \rho_0) G_1 + \rho_0 G_2$,  despite starting from the same game, can exhibit drastically different dynamics. \\ The system evolves by agents updating their strategies from the reciprocal interactions through an asynchronous process. In each round, a focal player $f$ and a model player $m$ are selected randomly. Each plays its strategy against $k$ (chosen as $k=4$) randomly chosen players of the population collecting a total payoff $\Pi_f$, $\Pi_m$ respectively. A stochastic update process depending on the payoff difference drives the strategy update of the focal player: with probability given by \begin{equation}
    p = \frac{1}{1+\exp(-\omega(\Pi_m-\Pi_f))}
\end{equation}
where $\omega$ is the selection strength 
(chosen as $\omega = 1$), the focal player adopts the strategy of the model player. After each round the cooperation level is updated and thus the following round payoff matrix is defined via the equation:
\[
G(\rho) = (1-\rho)\, G_1 + \rho\, G_2 .
\]
The simulation steps are repeated until the system stabilizes and reaches equilibrium, either by collapsing into an absorbing state of full defection/cooperation, or by reaching a stable nontrivial level of cooperation.
The code used for simulation is available at \textit{https://github.com/GiacomoFrigerio/CoevChimeraGames}. \\

\begin{figure}[ht]
        \centering
         \begin{minipage}{0.5\textwidth}
    \centering
\includegraphics[width=0.72\textwidth]{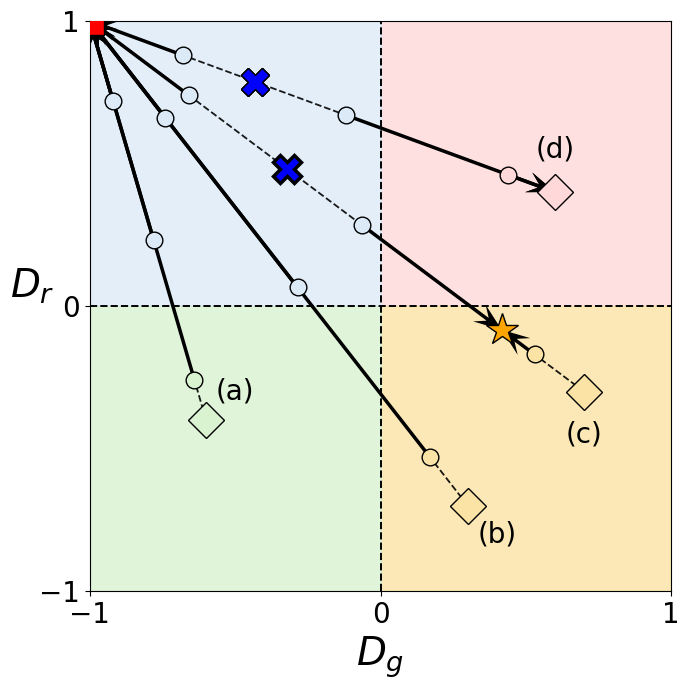}
\end{minipage}\hfill
  \begin{minipage}
    {0.5\textwidth}
\centering
\includegraphics[width=0.98\textwidth]{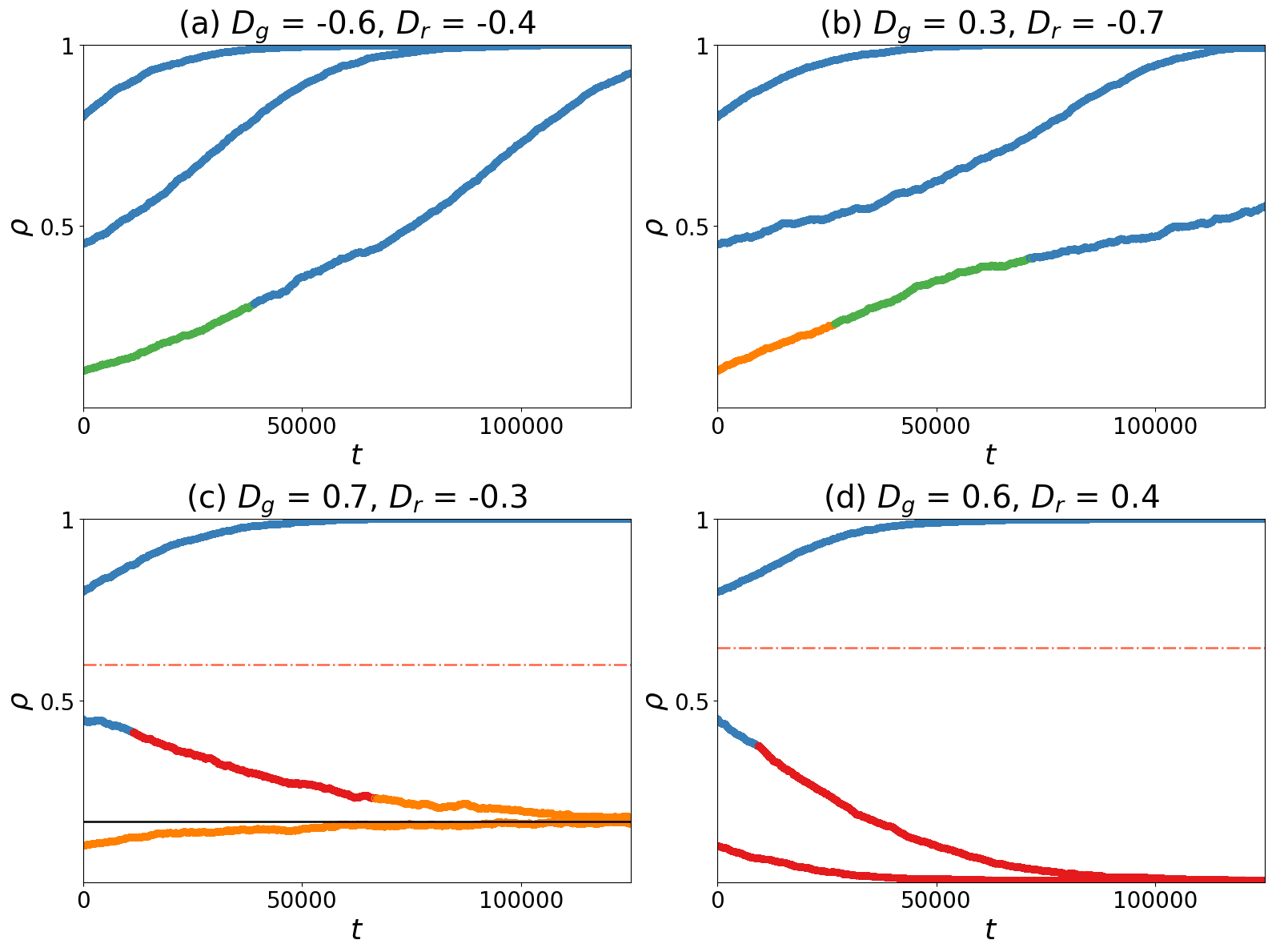}
  \end{minipage}
        \caption{\textbf{Dynamics of coevolutionary games with endogenous feedback with $G_2$ as a Stag Hunt.} The left panel shows the trajectory of the evolving game $G(t)$ within the social dilemma square as predicted analytically, while the right panels display the corresponding cooperation levels over time obtained via simulation. Throughout, $G_2$ is fixed as a Stag Hunt, and four different choices of $G_1$ are examined (diamonds).
        For each $G_1$, three distinct initial cooperation levels $\rho_0$ are considered, which determine three different initial games lying on the convex combination of $G_1$ and $G_2$ (circles). In the game square, colors indicate the location of nontrivial stable equilibria for feedback-driven games associated with each social dilemma class of $G_1$, thereby visualizing how endogenous feedback reshapes the parameter space at equilibrium. 
        Stable equilibria for the game played are highlighted as a star, unstable ones with a cross.
        In the simulation plots, the color of each trajectory highlights the social dilemma class of the instantaneous game $G(t)$.
        This scenario generally leads to absorbing states, an outcome coherent with the EGT predictions for a static Stag Hunt game. An interesting region lies in part of the SD class, where two solutions with opposite stability emerge: this is the only case of $\rho^\star \in (0,1)$ for certain $\rho_0$ values.
        }
        \label{fig:SHtarget:supp}
    \end{figure}

\begin{figure}[ht]
        \centering
         \begin{minipage}{0.5\textwidth}
    \centering
\includegraphics[width=0.72\textwidth]{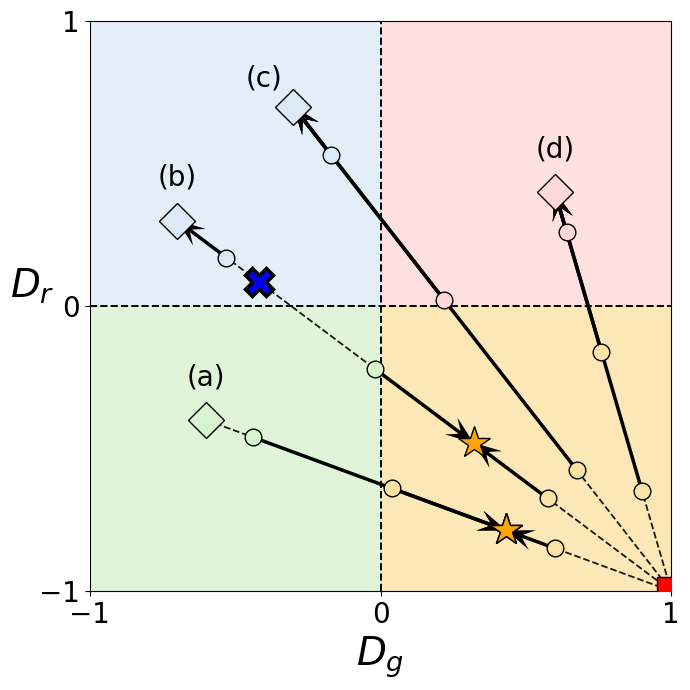}
\end{minipage}\hfill
  \begin{minipage}
    {0.5\textwidth}
\centering
\includegraphics[width=0.98\textwidth]{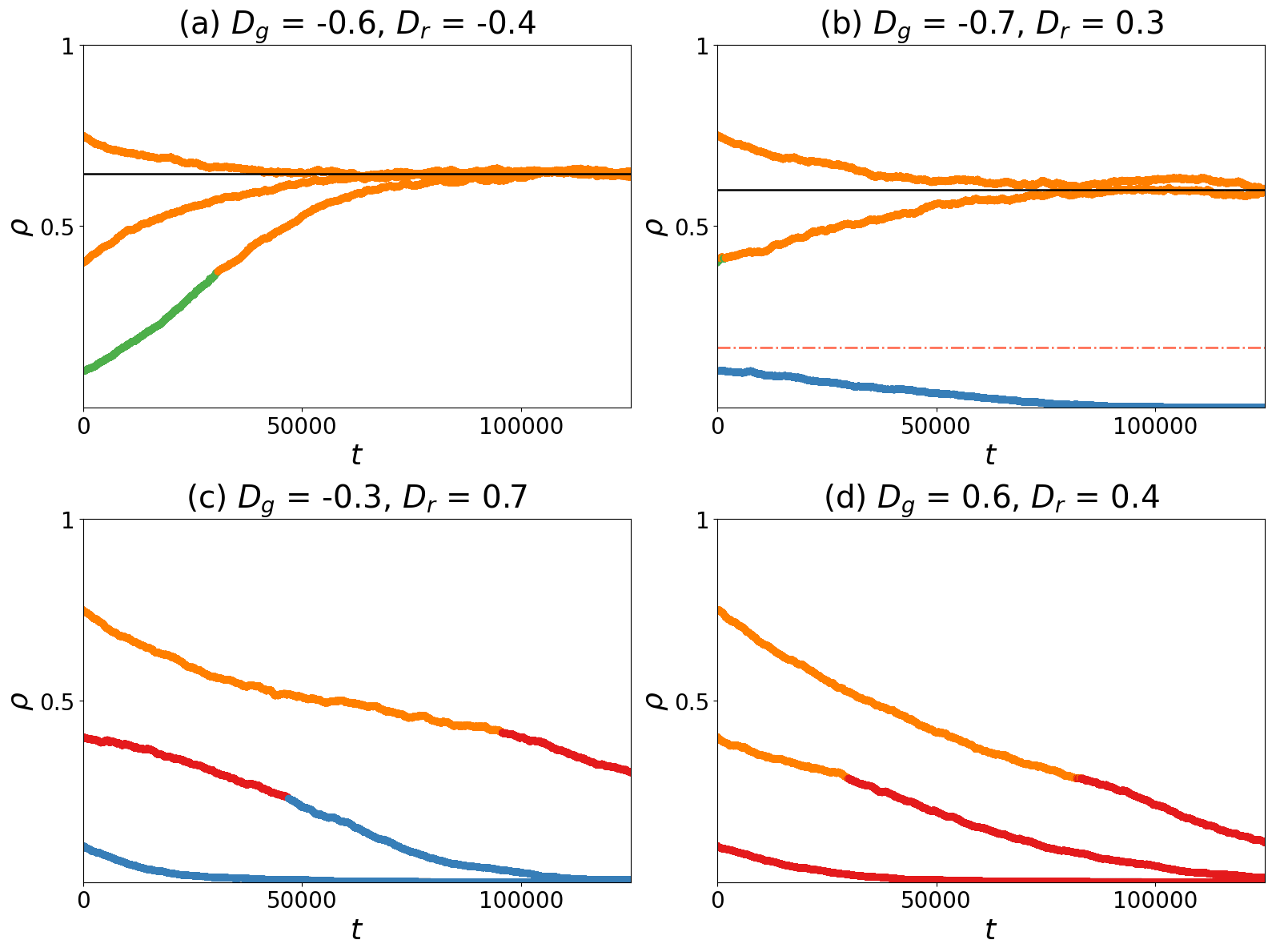}
\end{minipage}
        \caption{\textbf{Dynamics of coevolutionary games with endogenous feedback with $G_2$ as a Snowdrift Game.} The left panel shows the trajectory of the evolving game $G(t)$ within the social dilemma square as predicted analytically, while the right panels display the corresponding cooperation levels over time obtained via simulation. Throughout, $G_2$ is fixed as a Snowdrift game, and four different choices of $G_1$ are examined (diamonds).
        For each $G_1$, three distinct initial cooperation levels $\rho_0$ are considered, which determine three different initial games lying on the convex combination of $G_1$ and $G_2$ (circles). In the game square, colors indicate the location of nontrivial stable equilibria for feedback-driven games associated with each social dilemma class of $G_1$, thereby visualizing how endogenous feedback reshapes the parameter space at equilibrium. 
        Stable equilibria for the game played are highlighted as a star. Unstable equilibria are highlighted as a cross.
        In the simulation plots, the color of each trajectory highlights the social dilemma class of the instantaneous game $G(t)$.
        This scenario leads both to absorbing states ($G_1$ as PD, part of SH), and to nontrivial equilibria ($G_1$ as HG, part of SH). Symmetrically with the SH scenario, an interesting region lies in part of the SH class, where two solutions with opposite stability emerge.
}
        \label{fig:SDtarget:supp}
    \end{figure}

In Figures \ref{fig:PDtarget:supp}, \ref{fig:HGtarget:supp} below, we report the transformation in game square at equilibrium and the corresponding simulations performed in the PD/HG scenarios, which allow a single stable/unstable solution. In Figures \ref{fig:SHtarget:supp}, \ref{fig:SDtarget:supp}, analogous plots are shown for SH/SD scenarios. All plots clearly show the qualitative and quantitative agreement of the analytical results obtained via the replicator equation with the simulations.

\newpage

\section{Equilibria and stability}

Starting from the payoff matrix defined as 
\begin{equation}
     \bordermatrix {
 & C & D \cr
 C \ \ & R & S \cr
 D \ \ & T & P \cr} \ , \label{normal_game}
\end{equation}
 we consider a well-mixed population in which a fraction $\rho$ of individuals adopts cooperation and a fraction $1-\rho$ adopts defection. The expected payoffs accrued by cooperators and defectors when interacting randomly within the population are therefore
\[
\pi_C = \rho\, R + (1-\rho)\, S,
\qquad
\pi_D = \rho\, T + (1-\rho)\, P,
\]
where $R, S, T, P$ denote the standard payoff entries of the game introduced in Eq.~\ref{normal_game}. 

Within our co-evolutionary framework, however, these payoffs are not fixed: the payoff matrix evolves as a function of the cooperation level $\rho$ according to Eq.~\ref{game_matrix}. In particular, for the choice $f(\rho)=\rho$, the game played at state $\rho$ is the convex combination

\begin{equation}
\begin{split}
 G(t) & \equiv  G(\rho, D^1_g,D^1_r,D^2_g,D^2_r) = [1-f(\rho)]G_1 +f(\rho)G_2 
\\
& = [1-f(\rho)] \begin{pmatrix}
1 & -D^1_r \\
1+D^1_g & 0
\end{pmatrix} + f(\rho)
\begin{pmatrix}
1 & -D^2_r \\
1+D^2_g & 0
\end{pmatrix} \ . 
\end{split} \label{game_matrix}
\end{equation}
with $G_1$ and $G_2$ characterised by their respective dilemma strengths $D_r^i$ and $D_g^i$ ($i = 1,2$). Substituting this expression into the payoffs above yields a $\rho$-dependent payoff difference $\Delta \pi(\rho)=\pi_C(\rho)-\pi_D(\rho)$, which evaluates to
\begin{equation}
\label{generaldeltapayoff_supp}
\Delta \pi(\rho)
= \rho^{2}\bigl[(D_r^{2}-D_r^{1}) + (D_g^{1}-D_g^{2})\bigr]
\;+\;
\rho\,\bigl(2D_r^{1}-D_r^{2}-D_g^{1}\bigr)
\;-\;
D_r^{1}.
\end{equation}
This expression encapsulates the nonlinear feedback introduced by the co-evolution between the population state and the payoff structure. 
To determine the fixed points of the evolutionary dynamics, we consider the replicator equation
\begin{equation}
\label{replicator_supp}
\frac{d\rho}{dt}
= \rho(1-\rho)\,\Delta\pi(\rho),
\end{equation} 
In the long run, the population reaches an equilibrium characterized by a stable level of cooperation $\rho^\star$, while the game $G(t)$ converges to an 
\emph{equilibrium game} 
 $G_\mathrm{eq} = G(D_g^{\mathrm{eq}},D_r^{\mathrm{eq}}) =(1-\rho^\star)G_1+\rho^\star G_2$.
Two outcomes can occur: either the population reaches an absorbing state (i.e., $\rho^\star = 0$, full defection, or $\rho^\star = 1$, full cooperation), or it stabilizes on a level of cooperation $0 < \rho^\star < 1$, which correspond to the nontrivial solutions of $\Delta\pi(\rho)=0$. In the former case, at equilibrium the population plays $G_1$ or $G_2$, while in the latter case, it plays an equilibrium game $G_\mathrm{eq}$  different from $G_1$ and $G_2$: notably, we find that $G_\mathrm{eq}$ may even belong to a social dilemma class that is neither that of $G_1$ nor that of $G_2$.

The nontrivial levels of cooperation at equilibrium $\rho^\star$ can be determined by solving the RE in Eq.~\eqref{replicator_supp}. Specifically, setting $\Delta\Pi(\rho) = 0$ (with $f(\rho) = \rho$) yields: 
\begin{align}
\label{rho_star}
    \rho^\star = \frac{D_r^{\mathrm{eq}}}{D_r^{\mathrm{eq}}-D_g^{\mathrm{eq}}} 
    = \frac{D_r^1 +\rho^\star(D_r^2 - D_r^1)}{D_r^1 -D_g^1 +\rho^\star[(D_r^2 - D_r^1)-(D_g^2-D_g^1)]} \ .
\end{align}
The above equation admits at most two solutions for $\rho^\star \in (0,1)$, with opposite stability.
The choice of $G_1$ and $G_2$, combined with the aforementioned endogenous feedback in Eq.~(\ref{game_matrix}), gives rise to two distinct evolutionary outcomes, linking $\rho^\star$ to the game played at equilibrium $G_\mathrm{eq}$ in different ways. 
Either the population settles on cooperation levels that display stability properties consistent with those expected for the equilibrium game $G_\mathrm{eq}$ in the absence of feedback or it settles instead on a cooperative state $\rho^\star$ that 
subvert the nature of the equilibria expected for the standard (i.e., without endogenous feedback) evolutionary game 
dynamics for $G_\mathrm{eq}$. In the main text, we referred to this scenarios as \emph{chimera games} as they represent hybrid states in which a population reaches stable levels of cooperation that satisfy the Nash condition, but \emph{appear} to be inconsistent with respect to the game being played. In chimera games, equilibria that are unstable for standard evolutionary dynamics without feedback become stable, leading to novel and unexpected dynamical behaviors in social dilemmas.

To fully characterize our linear model, in the following we explore all possible pairs  \((G_1, G_2)\) 
of social dilemmas types. Each combination represents a distinct feedback configuration under which the population evolves, highlighting whenever chimera games emerge. We consider four different cases, according to the type of  social dilemma chosen for $G_2$. 

\begin{figure}[ht]
        \centering
         \begin{minipage}{0.5\textwidth}
    \centering
\includegraphics[width=0.72\textwidth]{lambda1PD_game.png}
\label{fig:PDtarget_gamesquare}
\end{minipage}\hfill
  \begin{minipage}
    {0.5\textwidth}
\centering
\includegraphics[width=0.98\textwidth]{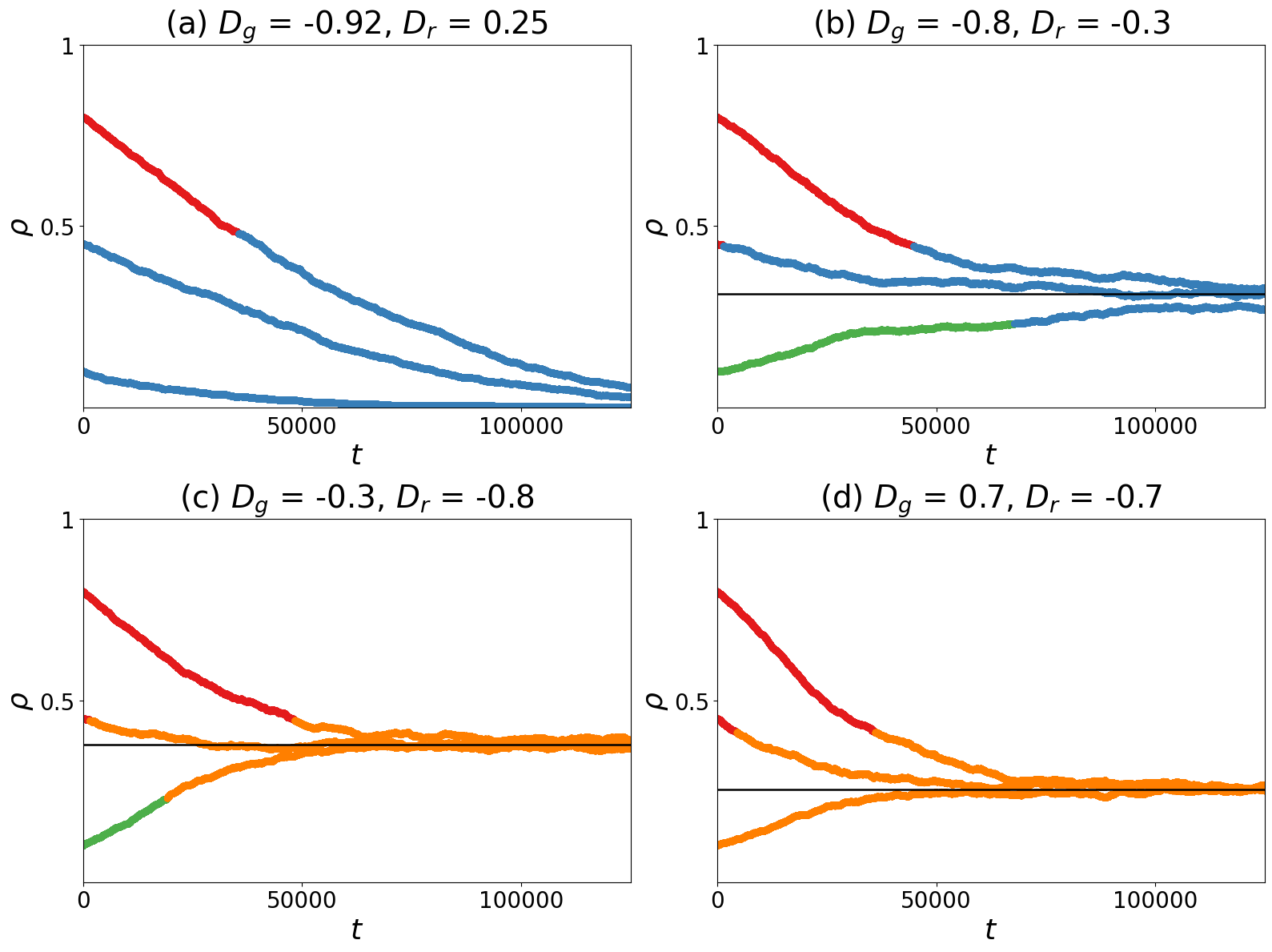}
\label{fig:PDtarget_simul}
  \end{minipage}
        \caption{\textbf{Dynamics of coevolutionary games with endogenous feedback with $G_2$ as a Prisoner's Dilemma.} The left panel shows the trajectory of the evolving game $G(t)$ within the social dilemma square as predicted analytically, while the right panels display the corresponding cooperation levels over time obtained via simulation. Throughout, $G_2$ is fixed as a Prisoner’s Dilemma, and four different choices of $G_1$ are examined (diamonds).
        For each $G_1$, three distinct initial cooperation levels $\rho_0$ are considered, which determine three different initial games lying on the convex combination of $G_1$ and $G_2$ (circles). In the game square, colors indicate the location of nontrivial stable equilibria for feedback-driven games associated with each social dilemma class of $G_1$, thereby visualizing how endogenous feedback reshapes the parameter space at equilibrium. 
        Stable equilibria for the game played are highlighted as a star.
        In the simulation plots, the color of each trajectory highlights the social dilemma class of the instantaneous game $G(t)$.
        As discussed in the main text, when $G_1$ is a Harmony Game (HG) with $D_r > D_g$, the dynamics invariably generate a chimera game, converging to a Stag Hunt equilibrium $G_{\mathrm{eq}}$ with a nontrivial fixed point $\rho^\ast$ (upper left). In contrast, for $G_1$ also in the HG class but with $D_r < D_g$, the expected EGT outcome arises from the Snowdrift equilibrium $G_{\mathrm{eq}}$ (upper right). A similar Snowdrift-type equilibrium is obtained when $G_1$ itself is a Snowdrift (SD) game (bottom left). Finally, if $G_1$ is a Stag Hunt (SH) game, the equilibrium game is again a SH, but with a unique absorbing state at $\rho^\ast = 0$ (bottom right).
}
        \label{fig:PDtarget:supp}
    \end{figure}

\begin{figure}[!ht]
    \centering

  \begin{minipage}{0.5\textwidth}
    \centering
\includegraphics[width=0.72\textwidth]{lambda1HG_game.png}
\label{fig:HGtarget_gamesquare}
\end{minipage}\hfill
  \begin{minipage}
    {0.5\textwidth}
\centering
\includegraphics[width=0.98\textwidth]{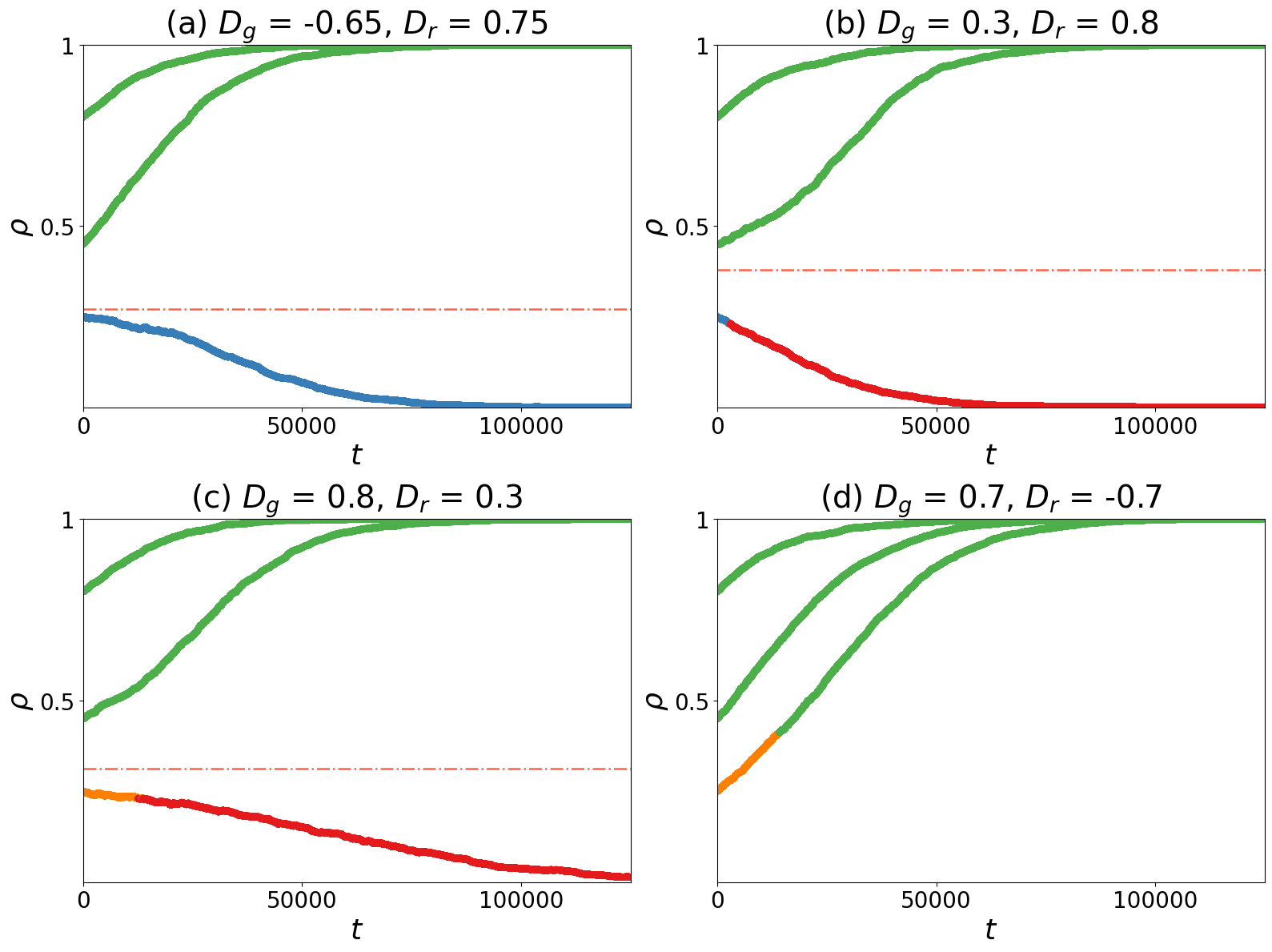}
\label{fig:HGtarget_simul}
  \end{minipage}
    \caption{\textbf{Dynamics of coevolutionary games with endogenous feedback with $G_2$ as a Harmony Game.} The left panel shows the trajectory of the evolving game $G(t)$ within the social dilemma square as predicted analytically, while the right panels display the corresponding cooperation levels over time obtained via simulation. Here, $G_2$ is fixed as a Harmony Game (HG), and several different choices of $G_1$ are examined (diamonds). For each $G_1$, three distinct initial cooperation levels $\rho_0$ are considered, which determine three different initial games lying on the convex combination of $G_1$ and $G_2$ (circles). In the game square, colors indicate the location of nontrivial (albeit unstable) equilibria for feedback-driven games associated with each social dilemma class of $G_1$, thereby visualizing how endogenous feedback reshapes the parameter space at equilibrium. 
    Unstable equilibria for the game played are highlighted as a cross.
    In the simulation plots on the right, the color of each trajectory highlights the social dilemma class of the instantaneous game $G(t)$. As discussed in the main text, and in contrast with the $G_2$ PD case, the nontrivial equilibria generated by endogenous feedback in this setting are always unstable. Consequently, the system can converge only to games corresponding to the original $G_1$ or $G_2$ payoff structures, and no chimera games arise in this regime.
}
   
    \label{fig:HGtarget:supp}
\end{figure}

\subsection{Prisoner's Dilemma as $G_2$}
\noindent When $G_2$ is a Prisoner's Dilemma, equation ~\ref{generaldeltapayoff_supp} becomes:
\begin{equation}
     \rho^2 (D_g^1 - D_r^1) + \rho (2D_r^1 - D_g^1 - 1) - D_r^1 = 0 \label{deltaPD_supp}
\end{equation}
By the intermediate value theorem,
at least a solution $\rho \in (0,1)$ exists in the region \( D_r^1 < 0 \): in fact, this choice gives $\Delta\Pi(0) = -D_r^1 >0$, while $\Delta\Pi(1) = -1 <0$. Due to the opposite value of $\Delta\Pi$ at the interval boundaries, and to the fact that the curve described by equation ~\ref{deltaPD_supp} has fixed concavity, 
it follows that this solution will also be unique.
The region of admissible nontrivial equilibria spans the Harmony Game (\( D_g^1 < 0 \)) and Snowdrift Game (\( D_g^1 > 0 \)) class. \\
Numerically, one finds that no physical solutions $\rho \in [0,1]$ are admissible for the regions where $D_r^1>0$, spanning the Prisoner's Dilemma (\( D_g^1 > 0 \)) and Stag Hunt (\( D_g^1 < 0 \)). 
Being $\Delta\Pi<0$ for any physical $\rho$, in both regions full defection is always the absorbing state reached by the system governed by Eq.~(\ref{replicator_supp}).
\\

\noindent Concerning the stability of the solutions found in the region $D_r^1 <0$, the Jacobian of the dynamical system $J(\rho) = \frac{\partial(d\rho/dt)}{\partial\rho}$, has to be studied at the stationary points; in fact, for our one dimensional case, it is simply the derivative. Using similar reasoning as before, it follows that a neighborhood of the root of Eq.~(\ref{deltaPD_supp}) is always negative, therefore the equilibrium is stable independently of $D_g^1$.
Regardless of the initial value of cooperation level $\rho_0$, the system will reach the cooperation level given by $\rho^\star$, solution of Eq.~(\ref{deltaPD_supp}).

\subsection{Harmony Game as $G_2$}
\noindent When $G_2$ is a Harmony Game, equation~\ref{generaldeltapayoff_supp} becomes:
\begin{equation}
     \label{deltaHG_supp}
     \rho^2 (D_g^1 - D_r^1) + \rho (2D_r^1 - D_g^1 + 1) - D_r^1 = 0 
\end{equation}

\noindent In this case we are guaranteed the existence of at least a solution $\rho \in (0,1)$ only in in the region \( D_r^1 > 0 \): $\Delta\Pi(0) = -D_r^1 <0$, while $\Delta\Pi(1) = 1 >0$. As before, such a solution will also be unique.
The region of admissible nontrivial equilibria spans the Stag Hunt (\( D_g^1 < 0 \)) and Prisoner's Dilemma (\( D_g^1 > 0 \)) class. \\
Numerically, one finds that no physical solutions $\rho \in [0,1]$ are admissible for the regions where $D_r^1<0$, spanning the Harmony Game (\( D_g^1 < 0 \)) and Snowdrift Game (\( D_g^1 > 0 \)).
Being $\Delta\Pi>0$ for any physical $\rho$, in both regions full cooperation is always the absorbing state reached by the system governed by Eq.~(\ref{replicator_supp}). \\

\noindent Concerning the stability of the solutions found in the region $D_r^1 >0$, using similar reasoning as before, it follows that a neighborhood of the root of Eq.~(\ref{deltaHG_supp}) is always positive, therefore the equilibrium is unstable independently of $D_g^1$.
As such, depending on the relative ordering between the initial value of cooperation level $\rho_0$ and the solution of Eq.~(\ref{deltaHG_supp}) $\rho^\star$, the system will either reach full cooperation ($\rho_0>\rho^\star$) or full defection ($\rho_0<\rho^\star$).


\subsection{Stag Hunt as $G_2$}
\noindent When $G_2$ is a Stag Hunt, equation~\ref{generaldeltapayoff_supp} becomes:
\begin{equation}
     \label{deltaSH_supp}
     \rho^2 (2+D_g^1 - D_r^1) + \rho (2D_r^1 - D_g^1 - 1) - D_r^1 = 0. 
\end{equation}
\noindent With similar arguments, we notice that in this case a unique solution $\rho \in (0,1)$ exists in in the region \( D_r^1 > 0 \): $\Delta\Pi(0) = -D_r^1 <0$, while $\Delta\Pi(1) = 1 >0$. This region of admissible nontrivial equilibria spans the Stag Hunt (\( D_g^1 < 0 \)) and Prisoner's Dilemma (\( D_g^1 > 0 \)) class. As for $G_2$ in HG, this solution is always unstable, resulting in bistability. \\
Contrarily to the previous cases, the region \( D_r^1 < 0 \) hosts a very rich solution landscape. 
To have real solutions, the determinant of Eq.~(\ref{deltaSH_supp}) must be greater than zero. The condition is the following:
\begin{equation}
    \label{determinant_SH}
    (D_g^1)^2+2D_g^1+4D_r^1+1 \geq 0.
\end{equation}
Points that do not satisfy this condition lie in part of the HG and also SD region, and display full cooperation due to $\Delta\Pi(\rho)>0$.
The remainder of these dilemma class regions hosts two distinct solutions $\rho^\star\in(0,1)$, of opposite stability (because of the fixed concavity of Eq.~(\ref{deltaSH_supp})), with the smaller one being stable: therefore, this region hosts a peculiar bistability, as $\rho_1^{\star}\in(0,1),\ \rho_2^{\star}=1$. \\
As mentioned in the main text, we can realize the nature of chimera games in the region delimited by the following conditions, spanning the HG and SD regions:
\begin{align}
D_r^1 &< 0 \\
(D_g^1)^2+2D_g^1+4D_r^1+1  &> 0 \\
-D_r^1 &>D_g^1,
\end{align}
because the resulting equilibrium game has $\rho^\star\in(0,1)$ while lying in the SH class.
Note that for $-D_r^1<D_g^1$ we have no chimera games as $G_\mathrm{eq}$ lies in the SD region and $\rho^\star\in(0,1)$ conforms to EGT predictions. 
As a last remark, we notice that technically there exists a degenerate line where equality for Eq.~(\ref{determinant_SH})  is realized, leading to a single nontrivial solution; however, this would be unattainable for any finite-size simulation and remains a technical nuance.

\subsection{Snowdrift as $G_2$}
\noindent When $G_2$ is a Snowdrift, equation~\ref{generaldeltapayoff_supp} becomes:
\begin{equation}
     \label{deltaSD_supp}
     \rho^2 (D_g^1 - D_r^1 - 2) + \rho (2D_r^1 - D_g^1 + 1) - D_r^1 = 0. 
\end{equation}
\noindent Due to symmetry, in this case of a unique solution $\rho \in (0,1)$ exists in in the opposite region than SH, \( D_r^1 < 0 \): $\Delta\Pi(0) = -D_r^1 >0$, while $\Delta\Pi(1) = -1 < 0$. This region of admissible nontrivial equilibria spans the Harmony game (\( D_g^1 < 0 \)) and Snowdrift (\( D_g^1 > 0 \)) class. As for $G_2$ in PD, this solution is always stable, leading to $G_\mathrm{eq}$ in the SD class. \\
Contrarily to the previous cases, the region \( D_r^1 < 0 \) hosts a very rich solution landscape. 
To have real solutions, the determinant of Eq.~(\ref{deltaSD_supp}) must be greater than zero. The condition is the following:
\begin{equation}
    \label{determinant_SD}
    (D_g^1)^2-2D_g^1-4D_r^1+1 \geq 0.
\end{equation}
Points that do not satisfy this condition lie in part of the PD and also SH region, and display full defection due to $\Delta\Pi(\rho)<0$.
The remainder of these regions hosts two distinct solutions $\rho^\star\in(0,1)$, of opposite stability (because of the fixed concavity of Eq.~\ref{deltaSD_supp}), with the larger one being stable: therefore, this region hosts a peculiar bistability, as $\rho_1^{\star}=0,\ \rho_2^{\star} \in(0,1)$. \\
As mentioned in the main text, no chimera games arise in this scenario, because for nontrivial equilibria $\rho^\star\in(0,1)$ the corresponding $G_\mathrm{eq}$ lies in the SD class.
Again, we notice the degenerate case where equality for Eq.~\ref{determinant_SD} is realized, leading to a single nontrivial solution.

    \newpage
\section{Effect of delay in the game}
\label{sec:delay_game}

We study how a finite response time in the incentive structure modifies the dynamics of endogenous coevolutionary games. 
To this end, we introduce a delay $\tau\ge 0$ in the order parameter that controls the game being played. 
In the delayed model, it is important to distinguish between the current population composition and the delayed state that determines the game. 
We therefore write
$$
x=\rho(t),
\qquad
y=\rho(t-\tau).
$$
The delayed replicator equation is
\begin{equation}
\dot{\rho}(t)
=
\rho(t)\bigl(1-\rho(t)\bigr)\,
\Delta\pi\!\bigl(\rho(t),\rho(t-\tau)\bigr),
\qquad \rho(t)\in[0,1],
\label{eq:delayed_replicator}
\end{equation}
where $\rho(t)$ denotes the current fraction of cooperators, while $\rho(t-\tau)$ determines the payoff matrix currently being played.

Delay differential equations in evolutionary dynamics have been investigated in several contexts, and recent work has emphasized ``social delay'' effects where payoffs are evaluated using past strategy frequencies while the underlying game remains fixed \cite{hu2023evolutionary,hu2026revisiting}. 
Here, instead, the delayed order parameter determines the \emph{game being played} through endogenous feedback. 
Thus, away from equilibrium, the delay generates genuinely time-dependent incentives.

For the linear mixing prescription, the game determined by the delayed cooperation level is
\begin{equation}
G(y)=(1-y)G_1+yG_2,
\qquad y=\rho(t-\tau),
\label{eq:delayed_linear_game}
\end{equation}
where
$$
G_i=
\begin{pmatrix}
1 & -D_r^{(i)}\\
1+D_g^{(i)} & 0
\end{pmatrix},
\qquad i=1,2.
$$
We define
$$
\Delta D_r=D_r^{(2)}-D_r^{(1)},
\qquad
\Delta D_g=D_g^{(2)}-D_g^{(1)}.
$$
The dilemma strengths at delayed state $y$ are therefore
\begin{equation}
D_r(y)=D_r^{(1)}+y\Delta D_r,
\qquad
D_g(y)=D_g^{(1)}+y\Delta D_g.
\label{eq:delayed_dilemma_strengths}
\end{equation}
At time $t$, however, the expected payoffs are computed using the current population composition $x=\rho(t)$. Hence
\begin{equation}
\pi_C(x,y)
=
x+(1-x)\bigl[-D_r(y)\bigr],
\end{equation}
and
\begin{equation}
\pi_D(x,y)
=
x\bigl[1+D_g(y)\bigr].
\end{equation}
The payoff difference is therefore
\begin{equation}
\Delta\pi(x,y)
=
\pi_C(x,y)-\pi_D(x,y)
=
-D_r(y)+x\bigl[D_r(y)-D_g(y)\bigr].
\label{eq:deltapi_xy}
\end{equation}
Equivalently,
\begin{equation}
\Delta\pi(x,y)
=
-\left(D_r^{(1)}+y\Delta D_r\right)
+
x\left[
D_r^{(1)}-D_g^{(1)}
+
y(\Delta D_r-\Delta D_g)
\right].
\label{eq:deltapi_xy_expanded}
\end{equation}

Observe that, in the delayed scenario, equilibria are unchanged by delay.
At a stationary state one has
$$
\rho(t)=\rho(t-\tau)=\rho^*.
$$
Therefore, the interior fixed-point condition is independent of $\tau$ and reads
\begin{equation}
\Delta\pi(\rho^*,\rho^*)=0.
\label{eq:interior_condition_delay}
\end{equation}
The absorbing states $\rho^*=0$ and $\rho^*=1$ are also fixed points because of the prefactor $\rho(1-\rho)$ in the replicator equation. 
Thus, delay does not alter the \emph{location} of equilibria; it affects only their stability.

For an interior equilibrium, substituting $x=y=\rho^*$ into Eq.~\eqref{eq:deltapi_xy} gives
\begin{equation}
-D_r(\rho^*)
+
\rho^*
\bigl[
D_r(\rho^*)-D_g(\rho^*)
\bigr]
=0.
\label{eq:interior_condition_explicit}
\end{equation}
Equivalently,
\begin{equation}
\rho^*
=
\frac{D_r(\rho^*)}
{D_r(\rho^*)-D_g(\rho^*)}.
\end{equation}
Expanding this condition gives the same quadratic equation as in the instantaneous linear-feedback case:
\begin{equation}
\rho^{*2}
\left[
\Delta D_r-\Delta D_g
\right]
+
\rho^*
\left[
D_r^{(1)}-D_g^{(1)}-\Delta D_r
\right]
-
D_r^{(1)}
=0.
\label{eq:rho_star_quadratic_delay}
\end{equation}
Equivalently,
\begin{equation}
\rho^{*2}
\left[
(D_r^{(2)}-D_r^{(1)})
+
(D_g^{(1)}-D_g^{(2)})
\right]
+
\rho^*
\left[
2D_r^{(1)}-D_r^{(2)}-D_g^{(1)}
\right]
-
D_r^{(1)}
=0.
\end{equation}

\paragraph*{Linear stability and Hopf-from-delay mechanism.}
Assume an interior equilibrium $\rho^*\in(0,1)$ exists and satisfies Eq.~\eqref{eq:interior_condition_delay}. 
Let
$$
\rho(t)=\rho^*+\eta(t),
\qquad
\rho(t-\tau)=\rho^*+\eta(t-\tau),
$$
with $|\eta|\ll 1$. 
Define
$$
F(x,y)=x(1-x)\Delta\pi(x,y).
$$
The delayed replicator equation can be written as
$$
\dot{\rho}(t)=F(\rho(t),\rho(t-\tau)).
$$
Linearising around $(\rho^*,\rho^*)$ gives
\begin{equation}
\dot{\eta}(t)
=
\alpha\,\eta(t)
+
\gamma\,\eta(t-\tau),
\label{eq:linear_dde_correct}
\end{equation}
where
\begin{equation}
\alpha
=
\left.
\frac{\partial F}{\partial x}
\right|_{(\rho^*,\rho^*)},
\qquad
\gamma
=
\left.
\frac{\partial F}{\partial y}
\right|_{(\rho^*,\rho^*)}.
\end{equation}
Since $\Delta\pi(\rho^*,\rho^*)=0$, the derivative of the prefactor $x(1-x)$ does not contribute at an interior equilibrium. Hence
\begin{equation}
\alpha
=
\rho^*(1-\rho^*)
\left.
\frac{\partial \Delta\pi}{\partial x}
\right|_{(\rho^*,\rho^*)},
\qquad
\gamma
=
\rho^*(1-\rho^*)
\left.
\frac{\partial \Delta\pi}{\partial y}
\right|_{(\rho^*,\rho^*)}.
\end{equation}
From Eq.~\eqref{eq:deltapi_xy}, one obtains
\begin{equation}
\frac{\partial \Delta\pi}{\partial x}
=
D_r(y)-D_g(y),
\end{equation}
and
\begin{equation}
\frac{\partial \Delta\pi}{\partial y}
=
-\Delta D_r
+
x(\Delta D_r-\Delta D_g).
\end{equation}
Therefore,
\begin{equation}
\alpha
=
\rho^*(1-\rho^*)
\left[
D_r^*-D_g^*
\right],
\label{eq:alpha_def}
\end{equation}
where
$$
D_r^*=D_r^{(1)}+\rho^*\Delta D_r,
\qquad
D_g^*=D_g^{(1)}+\rho^*\Delta D_g,
$$
and
\begin{equation}
\gamma
=
\rho^*(1-\rho^*)
\left[
-\Delta D_r
+
\rho^*(\Delta D_r-\Delta D_g)
\right].
\label{eq:gamma_def}
\end{equation}

Seeking exponential solutions $\eta(t)=e^{\lambda t}$ gives the characteristic equation
\begin{equation}
\lambda
=
\alpha+\gamma e^{-\lambda\tau}.
\label{eq:characteristic_correct}
\end{equation}
For $\tau=0$, this reduces to
$$
\lambda=\alpha+\gamma.
$$
Thus, the interior equilibrium is asymptotically stable in the instantaneous case if
\begin{equation}
\alpha+\gamma<0.
\end{equation}
Indeed, $\alpha+\gamma$ coincides with
$$
\rho^*(1-\rho^*)
\frac{d}{d\rho}
\Delta\pi(\rho,\rho)
\bigg|_{\rho=\rho^*},
$$
as expected from the instantaneous replicator dynamics.

To identify the onset of delay-induced oscillations, we set $\lambda=i\omega$ with $\omega>0$ in Eq.~\eqref{eq:characteristic_correct}. 
This gives
$$
i\omega
=
\alpha+\gamma\bigl[\cos(\omega\tau)-i\sin(\omega\tau)\bigr].
$$
Separating real and imaginary parts yields
\begin{equation}
0=\alpha+\gamma\cos(\omega\tau),
\label{eq:hopf_real}
\end{equation}
and
\begin{equation}
\omega=-\gamma\sin(\omega\tau).
\label{eq:hopf_imag}
\end{equation}
A Hopf crossing can occur only if
\begin{equation}
\gamma^2>\alpha^2.
\label{eq:hopf_condition}
\end{equation}
In this case,
\begin{equation}
\omega_c=\sqrt{\gamma^2-\alpha^2}.
\label{eq:omega_c_correct}
\end{equation}
The critical delay is
\begin{equation}
\tau_c
=
\frac{\theta_c}{\sqrt{\gamma^2-\alpha^2}},
\label{eq:tau_c_general}
\end{equation}
where $\theta_c$ is the smallest positive angle satisfying
\begin{equation}
\cos\theta_c=-\frac{\alpha}{\gamma},
\qquad
\sin\theta_c=-\frac{\sqrt{\gamma^2-\alpha^2}}{\gamma}.
\label{eq:theta_conditions}
\end{equation}
In the common case $\gamma<0$, this can be written as
\begin{equation}
\tau_c
=
\frac{1}{\sqrt{\gamma^2-\alpha^2}}
\arccos\left(-\frac{\alpha}{\gamma}\right).
\label{eq:tau_c_rep}
\end{equation}
For $\tau<\tau_c$ the interior equilibrium remains stable, while at $\tau=\tau_c$ a pair of complex conjugate eigenvalues crosses the imaginary axis, giving rise to a delay-induced Hopf bifurcation. 
For $\tau>\tau_c$, the equilibrium loses stability and the dynamics may approach a stable periodic orbit.

\begin{figure}[t]
    \centering
    \includegraphics[width=0.62\linewidth]{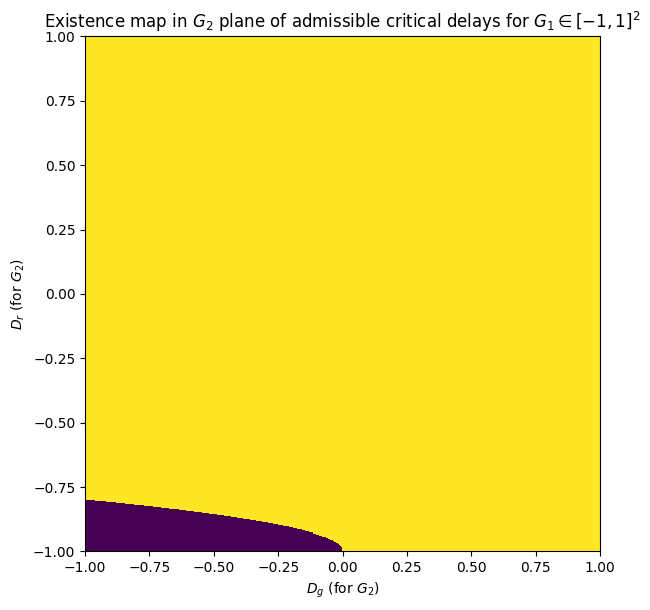}
    \caption{\textbf{Existence of delay-induced oscillations across the $G_2$ parameter space.}
    For each $G_2=G(D_g^2,D_r^2)$, we scan $G_1$ and determine whether there exists at least one $G_1$ such that the undelayed dynamics admits a stable interior equilibrium $\rho^*\in(0,1)$ and the delay can destabilize it through the Hopf mechanism described above. 
    Yellow denotes the existence of at least one such $G_1$, while purple indicates that no choice of $G_1$ in the scanned range produces a stable interior equilibrium satisfying the Hopf condition.}
    \label{fig:existence_G2}
\end{figure}

\subsection{Dependence on the choice of games}

We next analyse how the delay-induced oscillatory regime depends on the pair of games $(G_1,G_2)$. 
We first scan the $G_2$ parameter space and ask whether there exists at least one $G_1$ such that the undelayed dynamics admits a stable interior equilibrium $\rho^*\in(0,1)$ satisfying the delay-instability condition \eqref{eq:hopf_condition}. 
This is necessary for a Hopf-from-delay scenario. 
The resulting existence map is shown in Fig.~\ref{fig:existence_G2}.

To clarify the mechanism, we fix
$$
G_2:G(D_g,D_r)=G(1,1),
$$
and explore the parameter space of $G_1$. 
In this configuration, a delay-induced Hopf bifurcation can occur only if the undelayed dynamics $(\tau=0)$ admits a stable interior equilibrium $\rho^*\in(0,1)$, i.e. if Eq.~\eqref{eq:rho_star_quadratic_delay} admits an interior root with
$$
\alpha+\gamma<0.
$$
In addition, the delay term must be sufficiently strong relative to the instantaneous term, namely
$$
\gamma^2>\alpha^2.
$$
As discussed in the main text, stable interior equilibria for this $G_2$ are admitted when $G_1$ lies in the HG or SD region.

Within this feasible region, the critical delay in Eq.~\eqref{eq:tau_c_rep} varies smoothly across parameter space. 
As the boundary of feasibility is approached, either the undelayed equilibrium becomes weakly stable or the Hopf condition ceases to hold, and increasingly large delays are required to destabilise the equilibrium. 
Conversely, deeper inside the feasible region, smaller delays can trigger oscillatory behavior.

\begin{figure}[t]
    \centering
    \includegraphics[width=0.5\linewidth]{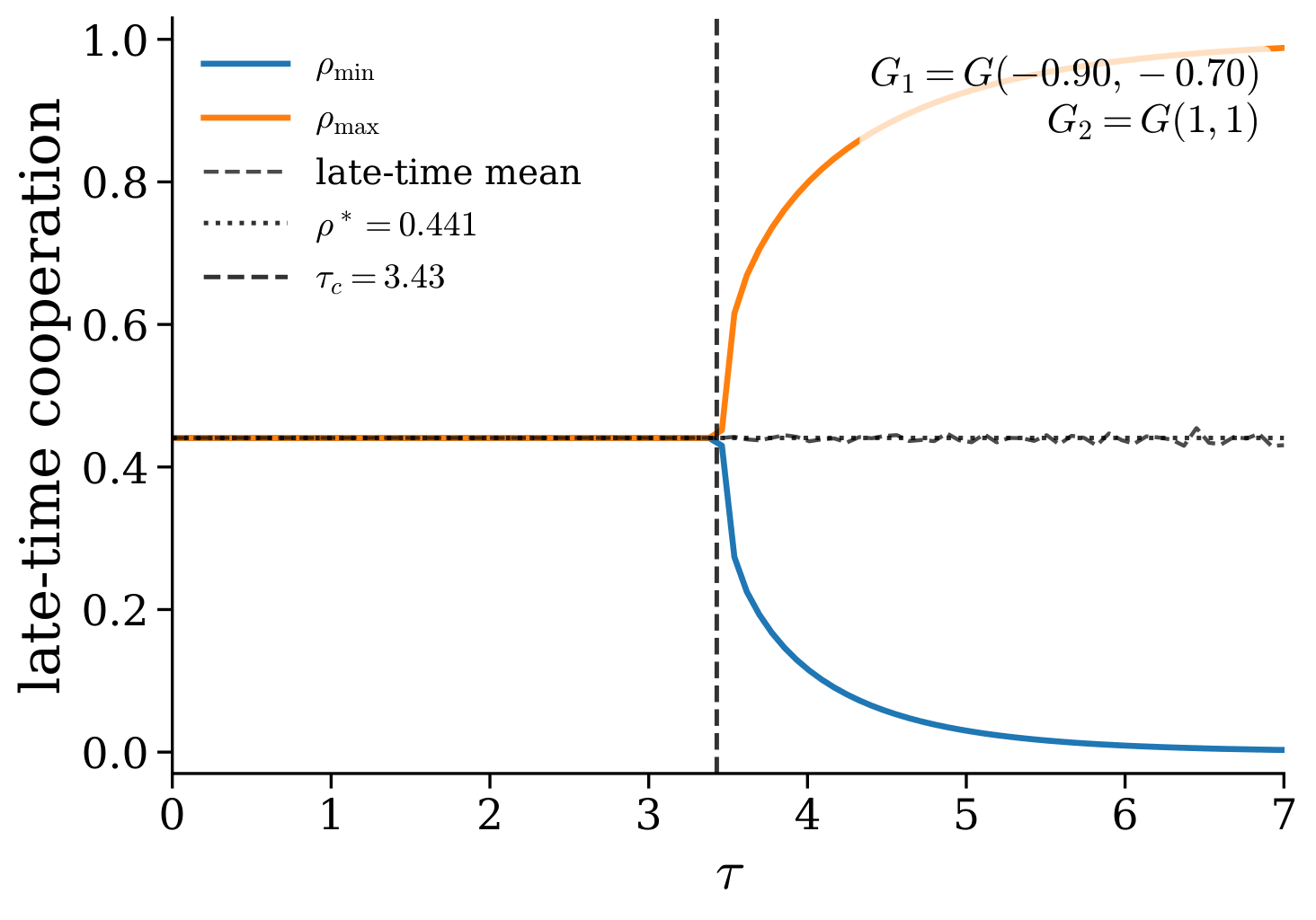}
    \caption{\textbf{Hopf bifurcation diagram for the delayed replicator dynamics.}
    For the representative pair $G_1:G(D_g,D_r)=G(-0.9,-0.7)$ and $G_2:G(D_g,D_r)=G(1,1)$, we plot the equilibrium branch $\rho=\rho^*$ together with the oscillation envelope $\rho_{\min}(\tau)$ and $\rho_{\max}(\tau)$ obtained from numerical simulations after discarding transients. 
    The vertical marker indicates the analytically predicted Hopf threshold $\tau_c\simeq 3.43$.}
    \label{fig:hopf_envelope}
\end{figure}

\begin{figure}[t]
\centering

\begin{minipage}{0.60\linewidth}
\centering
\includegraphics[width=\linewidth]{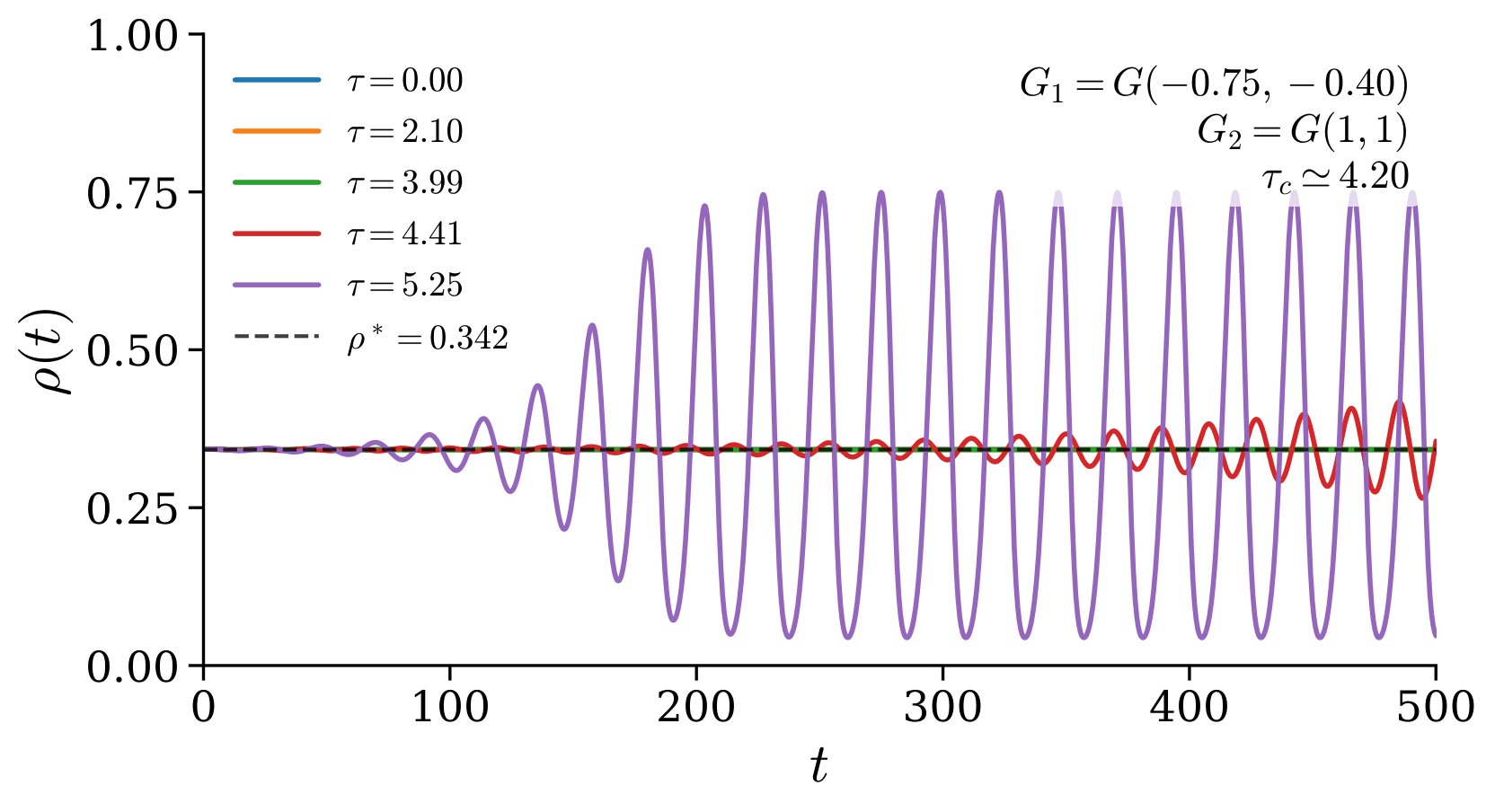}
\captionof{figure}{\textbf{Time series across the Hopf threshold.}
Delayed replicator trajectories $\rho(t)$ for the representative pair
$G_1:G(D_g,D_r)=G(-0.75,-0.4)$ and $G_2:G(D_g,D_r)=G(1,1)$, shown for increasing values of the delay $\tau$. 
For this pair the corrected analytical threshold is $\tau_c\simeq 4.20$. 
For $\tau<\tau_c$ trajectories relax to the coexistence equilibrium $\rho^*$, while for $\tau>\tau_c$ sustained oscillations emerge and grow in amplitude as the delay increases.}
\label{fig:timeseries_tau}
\end{minipage}
\hfill
\begin{minipage}{0.36\linewidth}
\centering
\includegraphics[width=\linewidth]{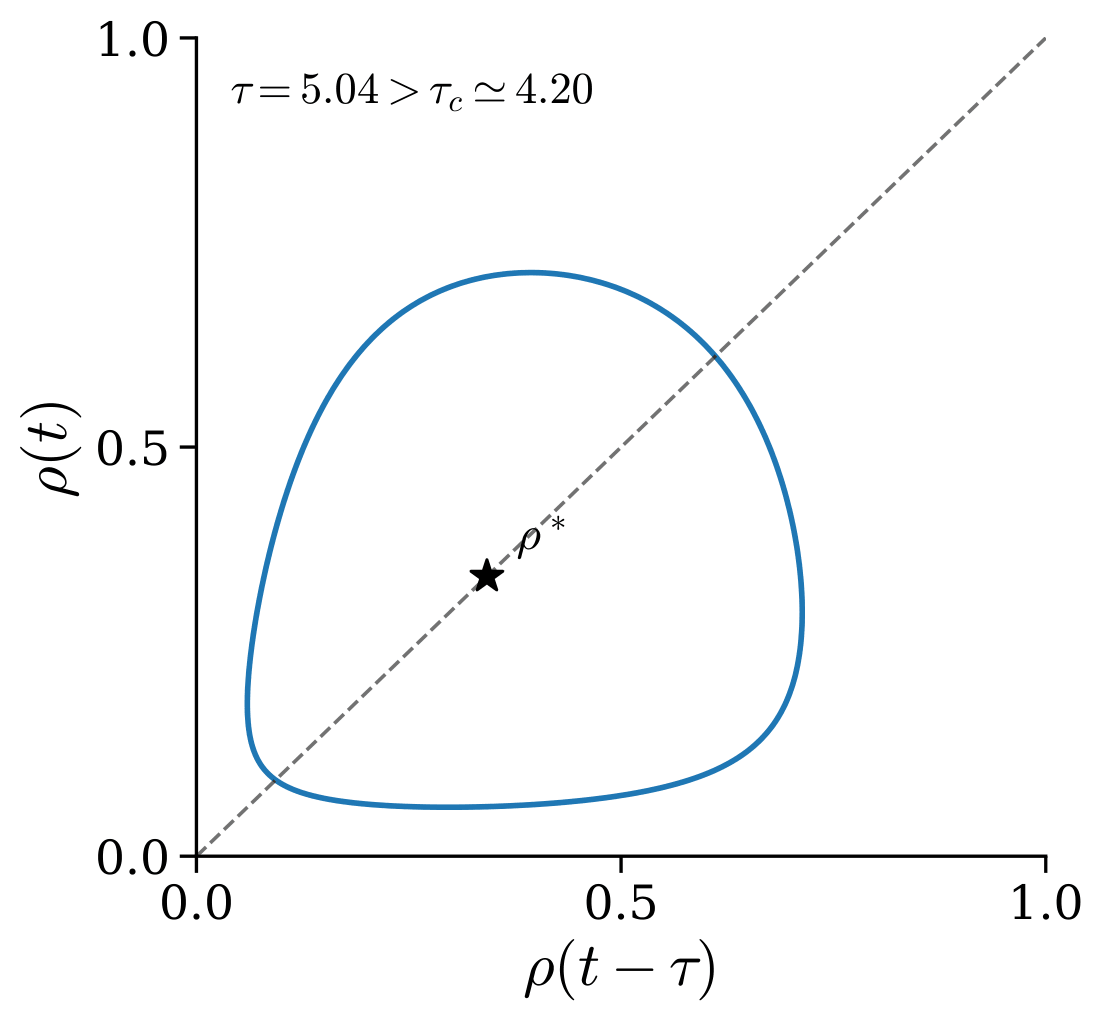}
\captionof{figure}{\textbf{Delay embedding of the oscillatory dynamics.}
Phase portrait in the delay-embedding plane $(\rho(t-\tau),\rho(t))$ for the representative pair shown in Fig.~\ref{fig:timeseries_tau} and for $\tau>\tau_c$. 
The closed loop indicates convergence to a stable limit cycle. 
The deviation from the diagonal $\rho(t)=\rho(t-\tau)$ quantifies the phase lag introduced by delayed feedback.}
\label{fig:embedding_delay}
\end{minipage}

\end{figure}

Let us now discuss in more detail a specific example. 
Consider the representative pair
\begin{equation}
G_1:G(D_g,D_r)=G(-0.9,-0.7),
\qquad
G_2:G(D_g,D_r)=G(1,1).
\label{eq:game_example}
\end{equation}
The undelayed dynamics admits a stable coexistence equilibrium
$$
\rho^*\simeq 0.441.
$$
For this equilibrium one finds
$$
\alpha\simeq 0.0276,
\qquad
\gamma\simeq -0.4408,
$$
so that
$$
\alpha+\gamma\simeq -0.4132<0,
$$
confirming stability at $\tau=0$. 
The Hopf condition $\gamma^2>\alpha^2$ is satisfied, and Eq.~\eqref{eq:tau_c_rep} yields
$$
\tau_c\simeq 3.43.
$$
Numerical simulations confirm that trajectories converge to $\rho^*$ for $\tau<\tau_c$ and approach a stable periodic orbit for $\tau>\tau_c$, with an oscillation amplitude that grows with $\tau$ (Fig.~\ref{fig:hopf_envelope}). 
The delay embedding $(\rho(t-\tau),\rho(t))$ provides a geometric signature of the oscillatory regime: below $\tau_c$ trajectories converge to $(\rho^*,\rho^*)$ on the diagonal, whereas above $\tau_c$ a closed loop appears, corresponding to a stable limit cycle (Fig.~\ref{fig:embedding_delay}).

\subsection{Bridging Monte--Carlo dynamics and delayed replicator theory}
\label{sec:bridge_mc_replicator}

We connect the critical delay derived from Eq.~\eqref{eq:delayed_replicator} to the discrete delay implemented in stochastic well-mixed simulations based on pairwise comparison with a Fermi update.

In the Monte--Carlo implementation, payoffs are evaluated using the delayed game
\begin{equation}
G\big(\rho(t-\tau)\big)
=
\bigl(1-\rho(t-\tau)\bigr)G_1
+
\rho(t-\tau)G_2,
\label{eq:delayed_game}
\end{equation}
and strategy updates follow
\begin{equation}
p_{i\leftarrow j}
=
\frac{1}{1+\exp\bigl[-\beta(\Pi_j-\Pi_i)\bigr]}.
\label{eq:fermi}
\end{equation}
Here $\Pi_\ell$ denotes the total payoff accumulated by individual $\ell$ from $k$ interactions, and $\beta\ge 0$ is the selection intensity. 
Let $\Delta\Pi=\Pi_C-\Pi_D$ denote the payoff difference between a cooperator and a defector sampled under the current population composition and the delayed game.

The expected change in $\rho$ in one elementary update is
\begin{equation}
\mathbb{E}[\Delta\rho]
=
\frac{1}{N}
\Big(
\Pr\{i=D,j=C\}\,\mathbb{E}[p_{D\leftarrow C}]
-
\Pr\{i=C,j=D\}\,\mathbb{E}[p_{C\leftarrow D}]
\Big),
\label{eq:delta_rho_expect}
\end{equation}
because $\rho$ increases by $1/N$ if a defector becomes a cooperator, and decreases by $1/N$ if a cooperator becomes a defector. 
In a well-mixed population with current cooperation level $\rho(t)$,
\begin{equation}
\Pr\{i=D,j=C\}
=
(1-\rho(t))\rho(t),
\qquad
\Pr\{i=C,j=D\}
=
\rho(t)(1-\rho(t)).
\end{equation}
By symmetry of the Fermi function,
\begin{equation}
p_{D\leftarrow C}
=
\frac{1}{1+e^{-\beta\Delta\Pi}},
\qquad
p_{C\leftarrow D}
=
\frac{1}{1+e^{+\beta\Delta\Pi}},
\end{equation}
and therefore
\begin{equation}
p_{D\leftarrow C}-p_{C\leftarrow D}
=
\tanh\left(\frac{\beta\Delta\Pi}{2}\right).
\end{equation}
Hence
\begin{equation}
\mathbb{E}[\Delta\rho]
=
\frac{1}{N}\rho(t)(1-\rho(t))
\,
\mathbb{E}
\left[
\tanh\left(\frac{\beta\Delta\Pi}{2}\right)
\right].
\label{eq:mc_drift_exact}
\end{equation}

Under weak selection, $\tanh z\simeq z$, and the expected payoff difference accumulated over $k$ interactions satisfies
\begin{equation}
\mathbb{E}[\Delta\Pi]
\simeq
k\,\Delta\pi\bigl(\rho(t),\rho(t-\tau)\bigr).
\end{equation}
The current cooperation level $\rho(t)$ appears because encounters occur in the current population, whereas the delayed value $\rho(t-\tau)$ appears because it determines the game being played. 
Thus
\begin{equation}
\mathbb{E}[\Delta\rho]
\simeq
\frac{1}{N}
\rho(t)(1-\rho(t))
\left(\frac{\beta k}{2}\right)
\Delta\pi\bigl(\rho(t),\rho(t-\tau)\bigr).
\label{eq:mc_drift_approx}
\end{equation}
Defining macroscopic time by $t=\ell/N$, where $\ell$ is the number of elementary updates, gives $\Delta t=1/N$ per update and yields the mean-field drift
\begin{equation}
\dot{\rho}(t)
\simeq
s\,\rho(t)\bigl(1-\rho(t)\bigr)
\Delta\pi\bigl(\rho(t),\rho(t-\tau)\bigr),
\qquad
s:=\frac{\beta k}{2}.
\label{eq:scaled_replicator}
\end{equation}

Linearising Eq.~\eqref{eq:scaled_replicator} around an interior equilibrium gives
\begin{equation}
\dot{\eta}(t)
=
s\alpha\,\eta(t)
+
s\gamma\,\eta(t-\tau),
\end{equation}
with the same $\alpha$ and $\gamma$ defined in Eqs.~\eqref{eq:alpha_def}--\eqref{eq:gamma_def}. 
The characteristic equation is therefore
\begin{equation}
\lambda
=
s\alpha+s\gamma e^{-\lambda\tau}.
\end{equation}
Consequently, the Hopf threshold in macroscopic Monte--Carlo time rescales as
\begin{equation}
\tau_c^{(\mathrm{MC})}
=
\frac{1}{s}\tau_c^{(\mathrm{rep})}
=
\frac{2}{\beta k}\tau_c^{(\mathrm{rep})},
\label{eq:tau_mc_vs_rep}
\end{equation}
where $\tau_c^{(\mathrm{rep})}$ is the corrected deterministic threshold in Eq.~\eqref{eq:tau_c_rep}. 
If the delay is implemented as a discrete number $r$ of elementary updates, then
\begin{equation}
r_c
=
N\tau_c^{(\mathrm{MC})}
=
N\frac{2}{\beta k}\tau_c^{(\mathrm{rep})}.
\label{eq:rc_discrete}
\end{equation}
Because this mapping is asymptotic and local, quantitative discrepancies are expected at finite $N$, finite $\beta$, and finite $k$, as well as due to stochastic payoff-sampling noise. 
Nevertheless, it provides a principled link between the deterministic Hopf threshold and the onset of oscillations in stochastic simulations.

\begin{figure}[t]
    \centering
      \includegraphics[width=0.7\linewidth]{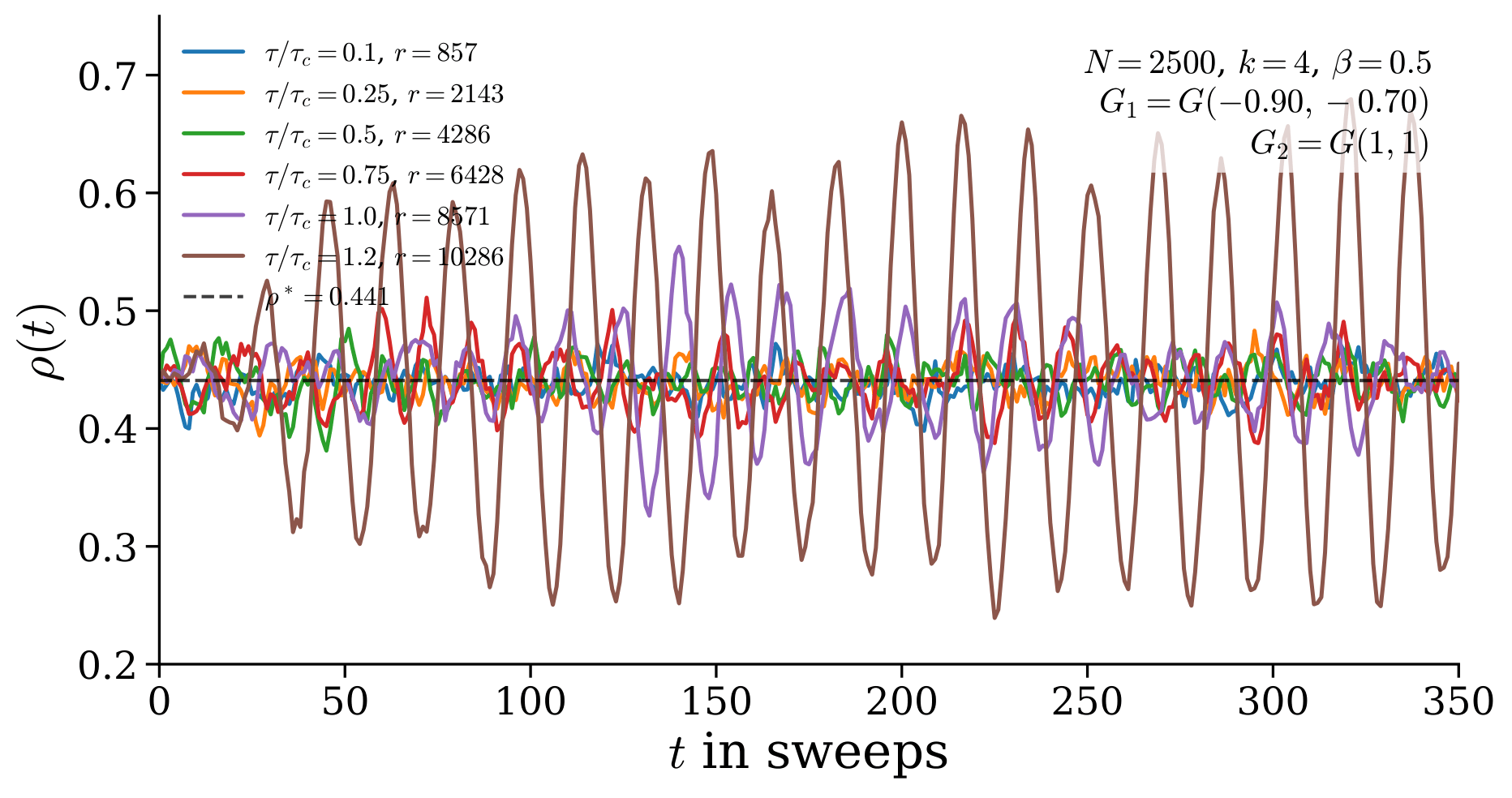}
    \caption{\textbf{Monte--Carlo trajectories for increasing delay at fixed selection intensity.}
    Time evolution of the cooperator fraction $\rho(t)$ in the well-mixed Monte--Carlo simulation at fixed selection intensity $\beta=0.5$ for the specific game pair in Eq.~\eqref{eq:game_example}. 
    Delays are expressed as fractions of the theoretical Hopf threshold, $\tau/\tau_c\in\{0.1,0.25,0.5,0.75,1.0,1.2\}$, where $\tau_c$ is computed using Eq.~\eqref{eq:tau_c_rep} and then rescaled according to Eq.~\eqref{eq:tau_mc_vs_rep}. 
    The legend reports the corresponding discrete delay $r$ in elementary updates used in the simulation. 
    For $\tau/\tau_c<1$ trajectories fluctuate around the coexistence level, whereas for $\tau/\tau_c\gtrsim 1$ oscillations persist with increasing amplitude, consistent with a delay-induced Hopf instability.}
    \label{fig:mc_timeseries_w05_taufrac}
\end{figure}

The MC simulation for the pair in Eq.~\eqref{eq:game_example} is shown in Fig.~\ref{fig:mc_timeseries_w05_taufrac}.

\clearpage
\section{Role of the exponent in the Feedback Function}

\begin{figure*}[t]
    \centering
    \includegraphics[width=\textwidth]{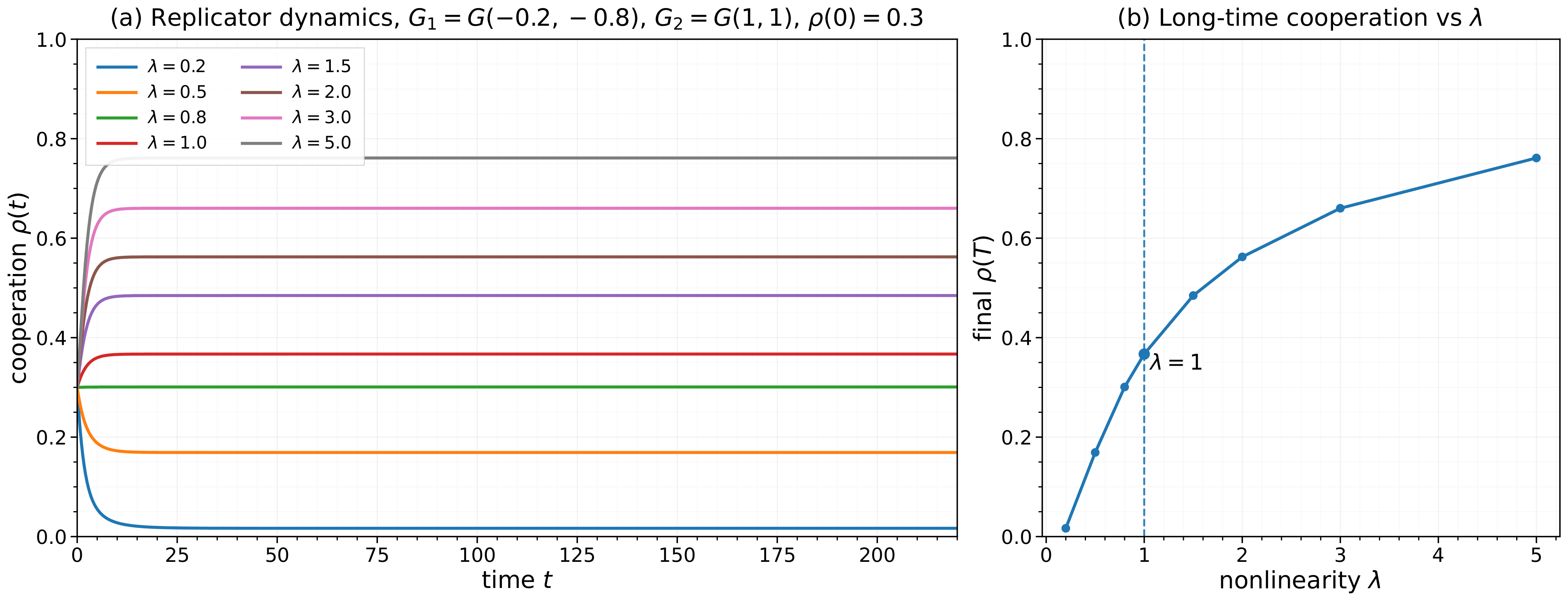}
    \caption{\textbf{Nonlinear endogenous feedback modulates cooperation under replicator dynamics.}
    We consider a population evolving under replicator dynamics,
    $\dot{\rho}=\rho(1-\rho)\Delta\pi(\rho)$, where the instantaneous game depends on the current cooperation level via
    $G(\rho)=(1-\rho^\lambda)G_1+\rho^\lambda G_2$.
    Here $G_2=G(D_g,D_r)=G(1,1)$ and $G_1=G(D_g,D_r)=G(-0.2,-0.8)$, with initial condition $\rho(0)=0.3$ (see panel title).
    \textbf{a} Time evolution $\rho(t)$ for different nonlinearities $\lambda$.
    \textbf{b} Long-time cooperation $\rho(T)$ at the end of the simulation window ($T=220$) as a function of $\lambda$; the dashed vertical line marks the linear case $\lambda=1$.
    Increasing $\lambda$ suppresses feedback at intermediate $\rho$ (since $\rho^\lambda<\rho$ for $\lambda>1$ and $\rho\in(0,1)$), shifting the asymptotic cooperation level.}
    \label{fig:lambda_sweep_replicator}
\end{figure*}

The linear specification $f(\rho)=\rho$, can be extended to the nonlinear mapping
\begin{equation}
f(\rho)=\rho^{\lambda},\qquad \lambda>0,
\end{equation}
so that the game played at cooperation level $\rho$ is
\begin{equation}
G(\rho)=\bigl[1-\rho^{\lambda}\bigr]\,G_1+\rho^{\lambda}\,G_2.
\label{eq:nonlinear_game_mapping_supp}
\end{equation}
This choice preserves the basic constraints discussed in the main text—continuity, monotonicity, and the boundary conditions $f(0)=0$, $f(1)=1$—while introducing a single parameter that controls the \emph{concavity} of the feedback. 
For $\lambda=1$ we recover the linear model. 
When $\lambda<1$, the mapping is concave and amplifies the influence of cooperation at intermediate frequencies ($\rho^{\lambda}>\rho$ for $\rho\in(0,1)$), representing a \emph{synergistic} feedback. 
Conversely, $\lambda>1$ produces a convex mapping and suppresses the contribution of $G_2$ until cooperation becomes sufficiently high ($\rho^{\lambda}<\rho$), capturing \emph{saturation} of incentive changes.

Introducing $\lambda\neq 1$ is therefore not a purely technical modification: it provides a parsimonious way to model cases in which incentives react disproportionately to early cooperative signals (synergy) or, instead, require widespread cooperation before substantially shifting (saturation). 
At the dynamical level, the nonlinearity reshapes the payoff difference $\Delta\pi(\rho)$ entering the replicator equation,
\begin{equation}
\dot{\rho}=\rho(1-\rho)\Delta\pi(\rho),
\end{equation}
and thereby affects both the location and the stability of interior equilibria, as well as the equilibrium game $G_{\mathrm{eq}}=G(\rho^\star)$ reached in the long run.
The payoff difference $\Delta\pi(\rho)$ appearing in the replicator equation 
is given, 
substituting $f(\rho) = \rho^\lambda$, by
\begin{equation}
   \Delta\pi(\rho) = \rho^{\lambda + 1}(D^2_r - D^1_r + D^1_g - D^2_g) + \rho^{\lambda}(D^1_r - D^2_r) + \rho (D^1_r - D^1_g) - D^1_r. 
\end{equation}
In particular, compared to the linear cases considered in the main text, the nonlinear mapping can shift the coexistence level $\rho^\star$ and modify which regions of the dilemma space are attainable as equilibrium games under endogenous coupling (see below and Supplementary Figures~\ref{fig:lambda_comparison}--\ref{fig:lambda_sweep_replicator}).
Increasing $\lambda$ progressively suppresses the feedback intensity at intermediate cooperation levels. 
Indeed, for $\lambda>1$ and $\rho\in(0,1)$ one has $\rho^\lambda<\rho$, so the contribution of $G_2$ to the game being played by the population is reduced unless cooperation becomes sufficiently large.

To illustrate the dynamical consequences of nonlinear endogenous feedback, we consider a representative pair of games with 
$G_1:G(D_g,D_r)=G(-0.2,-0.8)$ and $G_2:G(D_g,D_r)=G(1,1)$, and integrate the replicator dynamics starting from $\rho(0)=0.3$. 
Figure~\ref{fig:lambda_sweep_replicator} shows how the temporal evolution and the asymptotic level of cooperation depend on the nonlinearity parameter $\lambda$. 
Panel~(a) reports the trajectories $\rho(t)$ obtained for different values of $\lambda$, revealing that the population converges to different stationary cooperation levels depending on the strength of the nonlinear feedback. 
Panel~(b) summarizes this behavior by plotting the long-time cooperation $\rho(T)$, measured at the end of the simulation window, as a function of $\lambda$. 
The dashed vertical line marks the linear case $\lambda=1$, which corresponds to the linear feedback rule.

In the opposite configuration, where $G_2:G(D_g,D_r)=G(-1,-1)$ (a Harmony Game), the parameter $\lambda$ affects the equilibrium in a similar qualitative manner. As in the previous scenario, increasing $\lambda$ shifts the interior equilibrium toward higher cooperation levels. However, in this case the equilibrium $\rho^\star$ is unstable. 
Consequently, the nonlinear feedback modifies the position of this unstable threshold separating the basins of attraction of the absorbing states. For a population starting from an initial cooperation level $\rho_0$, smaller values of $\lambda$ lower the unstable equilibrium, effectively enlarging the basin of attraction of full cooperation. As a result, the system can cross the threshold and converge to the stable state $\rho^\star=1$.

The behavior of the solutions in two representative nonlinear regimes is now discussed, choosing $\lambda=2$ and $\lambda=0.5$ as paradigmatic examples of superlinear and sublinear feedback, respectively. 
For clarity, we focus on the main text case studies, where $G_2$ is a PD or HG, to highlight the role of the feedback mechanism.

\begin{figure}[!ht]
    \centering

  \begin{minipage}{0.48\textwidth}
    \centering
\includegraphics[width=0.9\textwidth]{lambda2PD_game.png}
\includegraphics[width=0.9\textwidth]{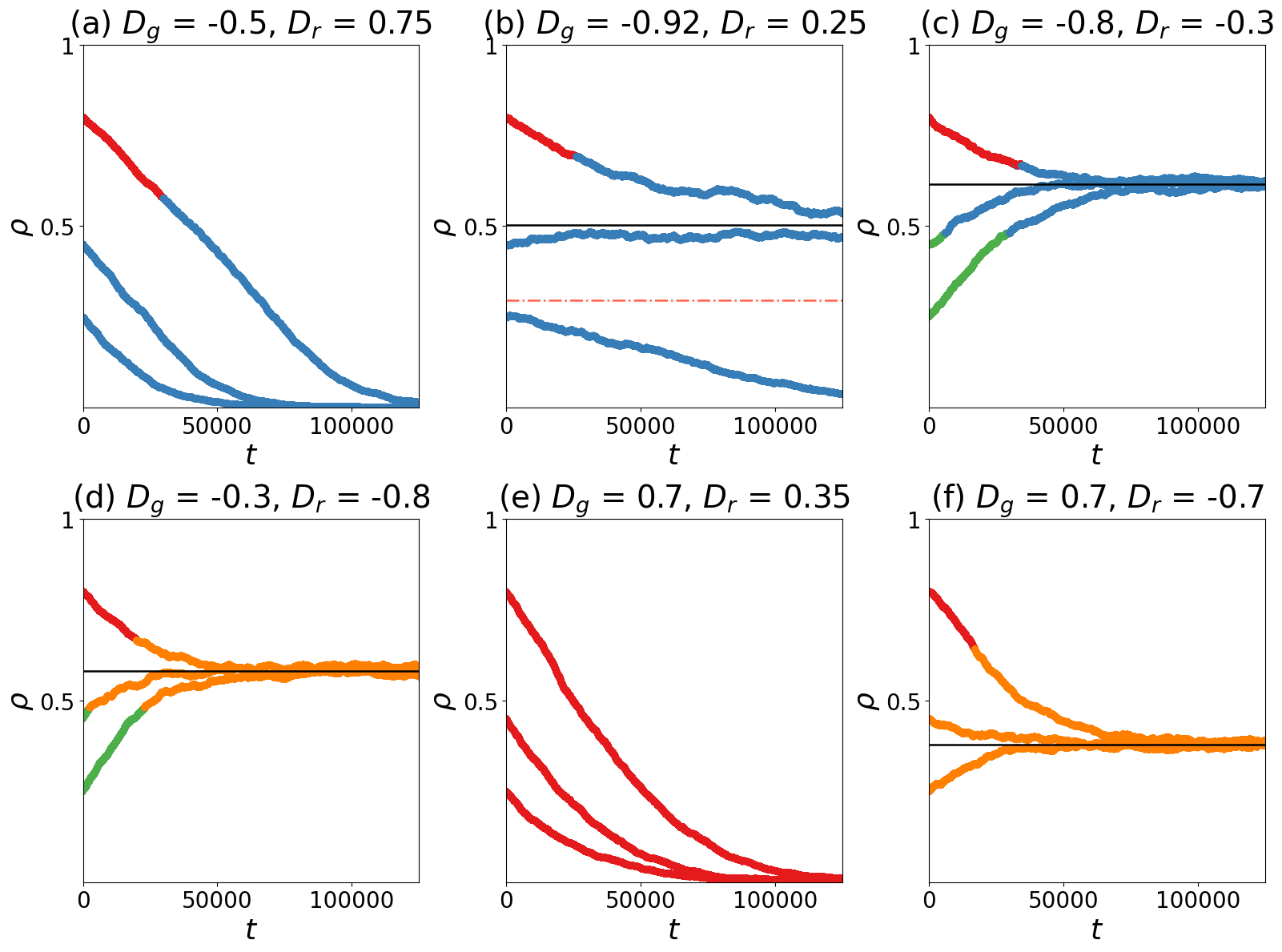}

\end{minipage}\hfill
  \begin{minipage}{0.48\textwidth}
\centering
\includegraphics[width=0.9\textwidth]{lambda05PD_game.png}
\includegraphics[width=0.9\textwidth]{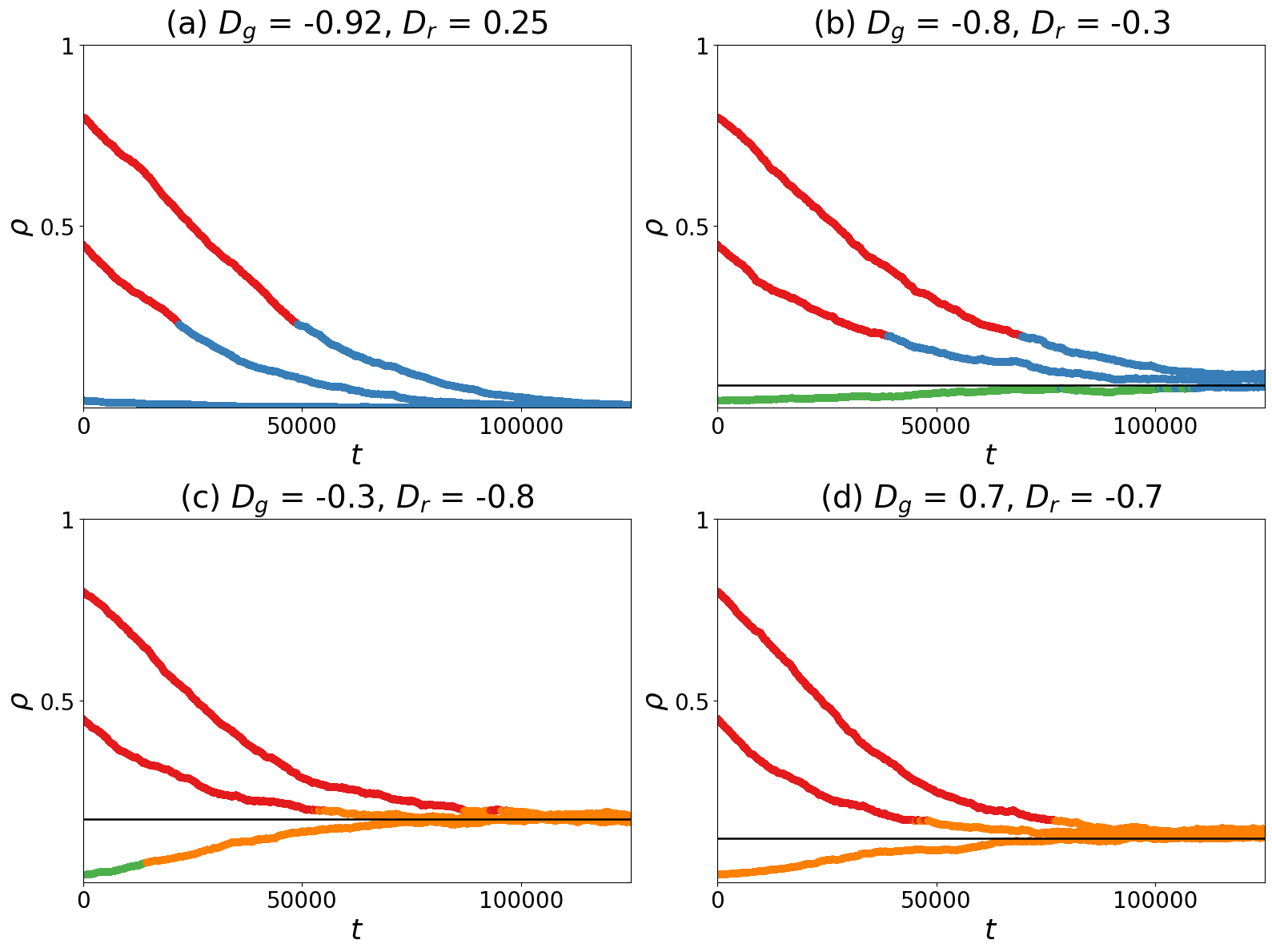}

  \end{minipage}
    \caption{\textbf{Evolution of games at equilibrium in the $(D^1_g,D^1_r)$ plane for different nonlinear feedback strengths, with $G_2$ fixed as a Prisoner’s Dilemma.}
    Diamonds indicate the chosen $G_1$ games. 
    For each $G_1$, three initial cooperation levels are considered, producing three initial games along the convex combination of $G_1$ and $G_2$ (circles). 
    Arrows show the trajectory from the initial game to the game reached at equilibrium under the endogenous feedback mapping $f(\rho)=\rho^\lambda$. 
    Stable equilibria are indicated by stars and unstable equilibria by crosses.
    \textit{Left: $\lambda=2$ (superlinear feedback).} Multiple arrows from the same initial condition indicate bistability: two equilibria may coexist (one stable, one unstable), both lying in the SH region but corresponding to different equilibrium cooperation levels.
    \textit{Right: $\lambda=0.5$ (sublinear feedback).} The qualitative equilibrium structure is similar to the linear case $\lambda=1$, but equilibrium cooperation levels are reduced due to the stronger sensitivity of the feedback.}
    \label{fig:lambda_comparison}
\end{figure}

\begin{figure}[!ht]
    \centering

  \begin{minipage}{0.48\textwidth}
    \centering
\includegraphics[width=0.9\textwidth]{lambda2HG_game.png}
\includegraphics[width=0.9\textwidth]{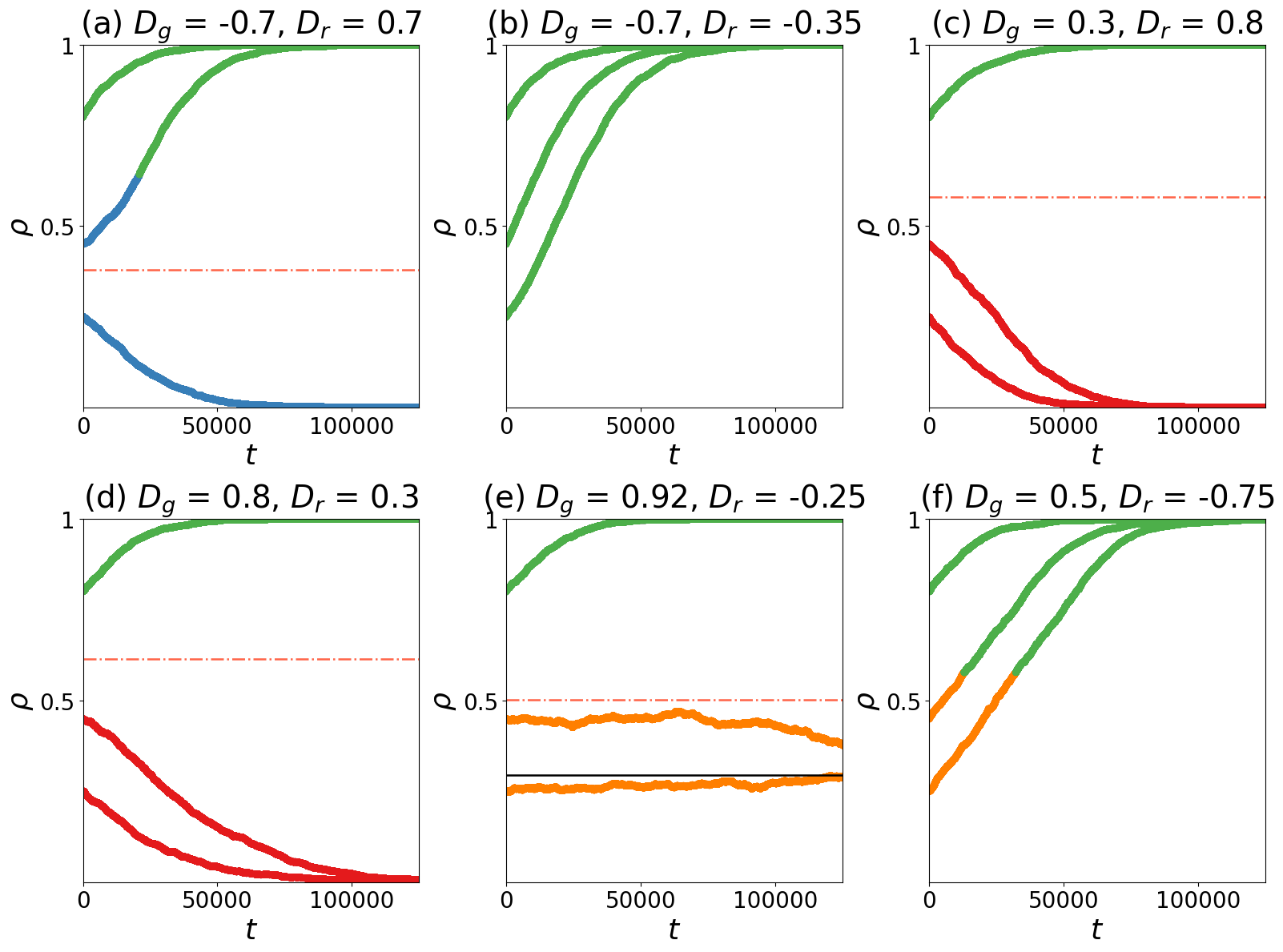}

\end{minipage}\hfill
  \begin{minipage}{0.48\textwidth}
\centering
\includegraphics[width=0.9\textwidth]{lambda05HG_game.png}
\includegraphics[width=0.9\textwidth]{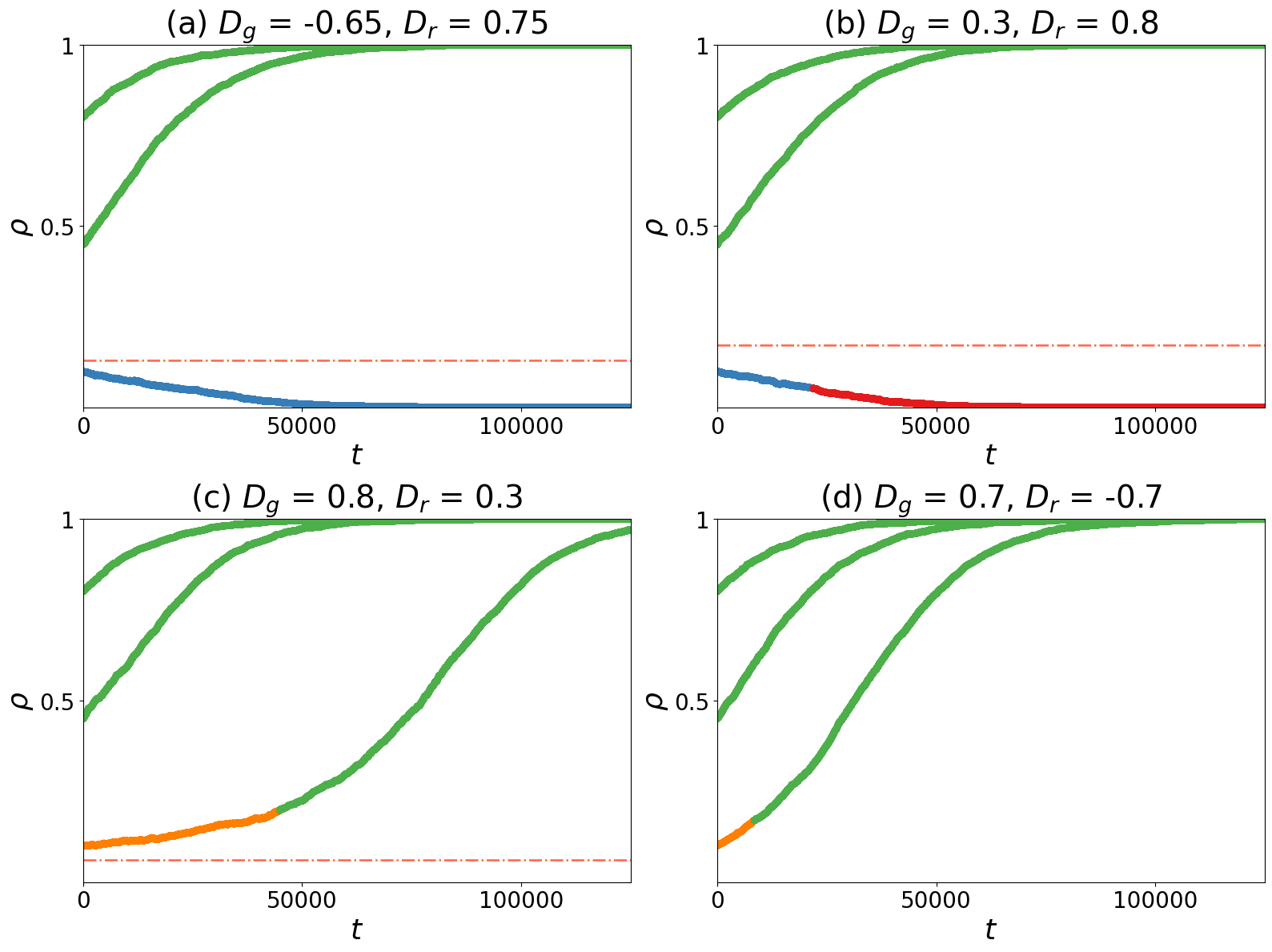}

  \end{minipage}
    \caption{\textbf{Evolution of games at equilibrium in the $(D^1_g,D^1_r)$ plane for different nonlinear feedback strengths, with $G_2$ fixed as a Harmony Game.}
    Diamonds indicate the chosen $G_1$ games. 
    For each $G_1$, three initial cooperation levels are considered, producing three initial games along the convex combination of $G_1$ and $G_2$ (circles). 
    Arrows show the trajectory from the initial game to the game reached at equilibrium under the endogenous feedback mapping $f(\rho)=\rho^\lambda$. 
    Stable equilibria are indicated by stars and unstable equilibria by crosses.
    \textit{Left: $\lambda=2$ (superlinear feedback).} Multiple arrows from the same initial condition indicate bistability: two equilibria may coexist (one stable, one unstable), both lying in the SD region but corresponding to different equilibrium cooperation levels.
    \textit{Right: $\lambda=0.5$ (sublinear feedback).} The qualitative equilibrium structure is similar to the linear case $\lambda=1$, but equilibrium cooperation levels are reduced due to the stronger sensitivity of the feedback.}
    \label{fig:lambda_comparison}
\end{figure}

\subsection*{Superlinear feedback}

We now consider superlinear feedback, for example $\lambda=2$, corresponding to $f(\rho)=\rho^2$. 
In most regions of the $(D_g^1,D_r^1)$ parameter space, the resulting dynamics qualitatively resembles the linear case $\lambda=1$. 
However, important qualitative deviations appear for specific combinations of the base games.
A particularly interesting scenario occurs when $G_1$ lies in the Stag Hunt (SH) region while $G_2$ is a PD. 
Under linear feedback, this configuration behaves effectively as a PD: defection is globally stable and no interior equilibrium exists. 
In contrast, when $\lambda=2$ the superlinear response of the feedback mechanism can generate two interior equilibria $\rho_1^\star$ and $\rho_2^\star$, with $\rho_1^\star<\rho_2^\star$, within the interval $(0,1)$. 
Both equilibria correspond to equilibrium games $G_{\mathrm{eq}}$ located in the SH region of the dilemma space.
Despite the SH classification of the equilibrium games, the resulting dynamics does not follow the classical SH scenario of evolutionary game theory, where only full cooperation or full defection can be stable outcomes due to bistability of the boundary equilibria. 
Instead, the system may stabilize at intermediate cooperation levels. 
These stationary states arise purely from the endogenous feedback between strategy frequencies and incentives, and cannot be inferred from the equilibrium game alone.
A similar mechanism appears when $G_1$ lies in the Snowdrift (SD) region while $G_2$ is a Harmony Game (HG). 
In the linear case, the cooperative nature of the HG would drive the population to full cooperation. 
With $\lambda=2$, however, the superlinear form of the feedback again allows two interior equilibria to emerge. 
Both correspond to games located in the SD region but sustain different cooperation levels, again with one stable and one unstable equilibrium.
In both examples, superlinear feedback fundamentally alters the dynamical structure of the system. 
Rather than simply shifting equilibrium points, it enables new coexistence states that do not appear in the linear model. 
These regimes illustrate a form of dynamical ambiguity: although the equilibrium game belongs to a well-defined class of social dilemmas, the actual population behavior does not correspond to the predictions of that game alone. 
This phenomenon exemplifies the chimera-game behavior discussed in the main text.

\subsection*{Sublinear feedback}

We now consider sublinear feedback, for example $\lambda=\frac{1}{2}$, corresponding to $f(\rho)=\rho^{1/2}$. 
In this regime the qualitative structure of the dynamics remains essentially the same as in the linear case $\lambda=1$: the same classes of equilibria appear across the parameter space and no additional coexistence states emerge.
The main effect of sublinear feedback is instead a quantitative shift of the equilibrium cooperation level. 
For instance, when $G_2$ is a PD the stationary cooperation level $\rho^\star$ is systematically lower than in the linear case. 
Because $\rho^{1/2}>\rho$ for $\rho\in(0,1)$, the feedback reacts more strongly to intermediate cooperation levels, driving the game being played more rapidly toward the defective configuration associated with $G_2$. 
As a result, cooperation is suppressed.
Importantly, the bistable interior regime observed for $\lambda=2$ in the SH--PD configuration disappears entirely. 
In this case the system again behaves effectively as a PD, with defection as the only stable outcome, mirroring the behavior of the linear model.

\bibliographystyle{apsrev4-2}
\bibliography{bibliography}%

\clearpage

\end{document}